\documentclass[a4paper,11pt]{article}
\pdfoutput=1 
\usepackage{jinstpub} 

 \usepackage{multirow}
 \usepackage{placeins}
 \usepackage{epstopdf}

\title{The optical system of the Tenerife Microwave Spectrometer: a window for observing the 10--20\,GHz sky spectra}

\author[a,b,1]{P. Alonso-Arias,\note{Corresponding author.}}
\author[a]{P.~A. Fuerte-Rodr\'{\i}guez,}
\author[a]{R.~J. Hoyland}
\author[a,b]{and J.~A. Rubi\~no-Mart\'{\i}n}

\affiliation[a]{Instituto de Astrofisica de Canarias,\\Calle V\'{\i}a L\'actea sn, ES38205 Santa Cruz de Tenerife, Spain}
\affiliation[b]{Universidad de La Laguna,\\Calle Padre Herrera sn, ES38205 Santa Cruz de Tenerife, Spain}

\emailAdd{pazaa@iac.es}

\abstract{ The TMS optical system is based on a decentered dual-reflector system in a Gregorian configuration to observe with an angular resolution of less than 2\textdegree.
  The primary goal of the present study is to evaluate the final design and verify that it satisfies the design requirements. We aim for low cross-polarization ($-$30\,dB), low sidelobe ($-$25\,dB) levels, and a stable beam in terms of shape (low ellipticity) and size over a full octave bandwidth (10--20\,GHz).
   We performed both ray-tracing and full-wave simulations using the CST Studio software in order to investigate the system behaviour. We  gave special attention to the beam frequency variation and polarization leakage. We have characterized the effects on the radiation pattern produced by the cryostat window.
   We present the final design of the TMS optical system, as well as a complete study of the system's performance in terms of cross-polarization, sidelobes, ellipticity and beamwidth. We discuss the effects of sidelobes and study the need for a baffle. }


\begin{document}
\maketitle
\flushbottom

\section{Introduction}\label{sec:intro}

Investigation of the CMB is fundamental to our understanding of the early history of the Universe. A major achievement in the study of the CMB was the COBE/FIRAS experiment \citep{fixen1994}, which measured its absolute temperature in the wavelength range from 1--95\,$\mathrm{cm^{-1}}$ with an angular resolution of 7\textdegree. These  results confirmed the blackbody form of the CMB spectrum, but lacked the resolution and sensitivity needed to detect the distortions that early energy release would have caused in the spectrum, as expected according to the $\Lambda$CDM Model (see \cite{sunyaev2013,chluba2016}). Since this result by COBE/FIRAS, absolute measurements of the CMB spectrum have not been further improved. This measurement gap is mainly due to the technological difficulty of obtaining measurements with sufficient accuracy ($\mathrm{\mu}$K variations) to detect the CMB spectral distortions and reveal interesting science.  The CMB community has been more focused in the characterization of the spatial distribution of this signal with space mission such as WMAP and Planck\footnote{\url{https://map.gsfc.nasa.gov/} and \url{http://www.esa.int/Planck}}. The study of spectral distortions would open a new unexplored window to the thermal history of the Universe and would allow us to explore the reionization epoch, the decaying and annihilating relics, or the recombination radiation, among many others (see \cite{chluba2019new} and references therein).

In the past years, absolute measurements of the CMB spectrum at longer wavelengths than COBE/FIRAS have been carried out with few ground-based and balloon-borne experiments (see eg. \cite{zannoni2008,singal2011}). Recently, several efforts have been made to overcome the technical difficulties of studying CMB spectral distortions at different wavelengths. New space missions like PIXIE \citep{kogut2011}, PRISM  \citep{andre2014}, PRISTINE\footnote{\url{https://www.ias.u-psud.fr/en/content/pristine}} or an spectro-polarimeter for ESA Voyage 2050 \citep{Delabrouille2019} have been proposed with the promise of detecting some of the above-mentioned spectral distortions. Unfortunately, none of them is approved at the moment, and if approved, their results will not be available earlier than 2030. In the meantime, ground-based experiments may allow exploring these spectral distortions, providing essential information for optimizing the design of future missions, either with new technological solutions or with key information on the astrophysical foreground emissions \citep{chluba2019new}. Moreover, an advantage of ground-based experiments is that they can potentially provide low frequency observations ($\mathrm{\nu\leq}$20\,GHz), which are harder to access with satellite missions. We would therefore be completing the spectral information on this signal for this frequency range. As a first step in this direction we present the Tenerife Microwave Spectrometer \citep[TMS, see][]{rubino2020}, which promises sufficient sensitivity in a band of great scientific interest to address some open questions. TMS is an ultra-high sensitivity radio-spectrometer that will observe the sky spectra in the 10--20\,GHz band at an altitude of 2400\,m from the Teide Observatory. 

The optical system of the TMS is the  window through which we will observe the sky spectra in the frequency range between 10--20\,GHz. It will gather and converge the light from a distance source in the sky, which travels as a free space wave, until it reaches the receiver. Common concerns regarding calibration and characterization of systematic errors due to optical systems in CMB experiments include: the control of sidelobes; the discrimination of orthogonal polarizations; and the control of the beam ellipticity and size variation throughout the operational bandwidth. These concerns affect the TMS experiment and its technical requirements in the same way: 1) the required level of sensitivity \citep{rubino2020} imposes an extremely low system temperature and, to ensure that no unwanted sources in the sky add to this temperature, we require a maximum sidelobe level of $-$25\,dB; 2) the pseudo-correlation architecture of the TMS radiometer \citep{alonsoarias2020} takes advantage of the use of both orthogonal polarizations to double its reception capacity, for which we require excellent cross-polarization discrimination, lower than $-$30\,dB; and 3) although it was not the original goal, the TMS has polarimetric capabilities \citep{rubino2020,alonsoarias2020}, and therefore, stabilizing the beam ellipticity around 4\% throughout the band is key to reduce errors when reconstructing the linear spectra. To meet all of these requirements, the TMS relies on a Gregorian optical configuration, with a 1.5\,m parabolic primary mirror, and a 0.6\,m elliptical secondary mirror.

In this work, we present the TMS optical design, including an overview of its main optical sub-systems and a complete performance analysis report. It is organized as follows: section~\ref{sec:overview} briefly summarises some formal aspects of the TMS, including the scan strategy, the general layout of the optical system and the description of the enclosure and telescope mount. In section~\ref{sec:design} we deal with the design of the optical system and the considerations we took into account in order to obtain optimal performance. We also include in this section the description and characterisation of each of its components, i.e., the mirrors, the feedhorn and the cryostat window. An exhaustive study of the system's  performance is presented in section~\ref{sec:preliminarstudy}, including an analysis of the effects of the cryostat window and the frequency and polarization dependence. This in-depth analysis of the beam response, in addition to accurate measurements during the comissioning phase, shall prove essential for the development of  systematics reduction and calibration software.

\section{Overview on the TMS experiment}\label{sec:overview}

\subsection{Scan Strategy of the TMS}\label{sec:scanstrategy}

TMS will scan the sky with a maximum azimuth velocity, or scan speed, of 20\textdegree $\mathrm{/s}$, and a minimum of 15\,arcsec/s. The mount guarantees a zenith distance from $-$10\textdegree~to 60\textdegree, with a maximum pointing error during tracking of 1\,arcmin in 60\,minutes.

TMS will conduct two types of surveys in order to achieve the scientific goals described in \cite{rubino2020}, with a similar sky coverage to that of the QUIJOTE Experiment. First, in the nominal mode, we will perform wide surveys at constant elevation of typically 60\textdegree. After two years of effective integration, a sensitivity of $\mathrm{10^{-25}\,W\,m^{-2}\,Hz^{-1}\,sr^{-1}}$ at 10\,GHz could be achieved in about 18200\,$\mathrm{deg^2}$. Additionally, dedicated raster scan observations will allow us to perform deep surveys in certain sky areas of interest, such as the area observed by ARCADE 2, \cite{singal2011}.

\subsection{General layout of the Optical System} \label{sec:layout}

The technical requirements needed to measure the absolute CMB spectrum are extremely demanding. In order to achieve the goals of the TMS project in terms of both sensitivity, angular resolution and systematics, a great technological effort was made. The TMS is designed to observe the sky with an angular resolution (FWHM)  of no more than 3\textdegree, to provide a discriminating factor between orthogonal polarizations of $-$30\,dB, and to control near sidelobes levels below $-$25\,dB in the whole operational band. In addition, beam size and shape must be analytically comparable over the band, so we are able to reconstruct the linear polarization spectra (Q and U Stokes parameters). Table~\ref{tab:specs} summarizes the basic optical characteristics of the TMS (see \cite{rubino2020, alonsoarias2020} for more information about the instrument technical requirements and design). 

\begin{table}
\caption{Optical system requirements for the TMS experiment. See text for details.} 
\label{tab:specs}
\begin{center}       
\begin{tabular}{lc}
\hline
Frequency range [GHz] & $10.0$--$20.0$ \\
Beam FWHM [deg] & 3.0 \\
Beam ellipticity & $\leq$10\,\%  \\
Cross-polarization (across band) & $< -$30\,dB\\
Sidelobe levels (across band)  & $<-$25\,dB \\
\hline
\end{tabular}
\end{center}
\end{table}

The optical system consists of two mirrors, a transparent-to-microwave-wavelengths cryostat window, a sky and reference feedhorn, a blackbody calibrator and other opto-mechanic components such as OrthoMode Transducers (OMTs) and 180\textdegree~hybrid couplers, and the structures and subsystems that support them. In this work, we focus on the optical configuration and its performance, in addition to the analysis of the effects of the cryostat window.

\subsection{Enclosure and telescope mount}\label{sec:enclosure}

The TMS instrument will be located in the Teide Observatory, at 2400\,m, in the former platform of the VSA experiment \citep{VSA1}, near the QUIJOTE experiment \citep{QUIJOTE_SPIE2012}. Both the telescope mount and the instrument will be integrated inside an oval fiberglass All-Sky dome of 4.5\,m with a height of 4.3\,m (see \cite{rubino2020}), which covers an observation area of the sky of 180\textdegree. Taking into account that the focal point of the instrument will be set at 1.8\,m above the ground, and considering the dimensions of the building, the instrument will be able to scan the sky in elevation down to 60\textdegree~with respect to the zenith. The structure, which will be attached to a concrete foundation, has an accessible entrance door to allow maintenance of both instrument and dome. A feedthrough located on one of the lateral walls allows the different wires and connections (electrical, compressed air, water and helium) to enter the dome. The dome includes several actuators and sensors to control its movement by means of a microcontroller, which will be connected to a server-logger to enable remote operation. In addition, a weather station outside the dome will monitor the environmental conditions at the observatory and allow an automatic evaluation and response to close the dome if necessary. The dome environment minimizes thermal fluctuations during the scientific exploitation stage.

Regarding the telescope mount, the TMS spectrometer shall be mounted on an alt-azimuth mount that is capable of keeping the instrument at a given declination angle while scanning through the azimuthal axis. Its design, which is based on the QUIJOTE telescopes \citep[see][]{QUIJOTE_SPIE2012}, consists of a base pedestal that allows an azimuth movement, and a fork. The instrument will be coupled to the fork, allowing elevation movement. This design guarantees continuous azimuth rotation and an elevation range from 0\textdegree~to 60\textdegree~with respect to the zenith. The mount will be remotely operated, with a minimum scan speed of 15\,$\mathrm{arcsec^{-1}}$ and a pointing RMS error of 1\,arcmin in 60 minutes. 

With the instrument installed on the platform, a dual-reflector system will focus the sky signal on the window of the cryostat. The secondary mirror (SM) connects directly to the instrument. This mirror can be removed to observe without optics, pointing the instrument directly at the sky. An absorber material (e.g. Eccosorb) shall be mounted around both primary and secondary mirrors, as well as the adjacent areas of the supporting structure,  to minimize unwanted multiple reflections in the dome. Oblique reflections on the conductive walls of the dome can lead to
higher levels of sidelobes which would increase the system noise, and to the presence of spurious polarized signals even when we are observing unpolarized sources (polarization leakage). The use of absorbing-coated structures ensures maximum acceptable levels of secondary lobes and the instrument induced polarization.

\section{The eye of the TMS: the optical system design}\label{sec:design}

\subsection{The TMS feedhorn}\label{sec:TMSfeed}

The TMS feedhorns have been designed and manufactured to ensure strict radiation requirements regarding cross-polarization (XPol) and return loss (RL). 
A novel design based on a metasurface pattern was used to provide XPol and RL levels below $-$35 and $-$25\,dB, respectively, over the operating bandwidth, between 10--20\,GHz. The  pattern was optimized with CST Studio Suite\footnote{\url{https://www.3ds.com/products-services/simulia/products/cst-studio-suite/}} and subsequently modified to make it implementable. This modification allowed the mechanical workshop at the IAC to manufacture the feedhorns as a set of aluminium rings implementing the metasurface (platelets technique). More detailed information about the design and manufacture processes of these meta-horns can be found in \cite{demiguel2019_hornfund} and \cite{demiguel2021metamaterial}.  

\begin{figure*}%
\centering
\includegraphics[width=0.45\textwidth]{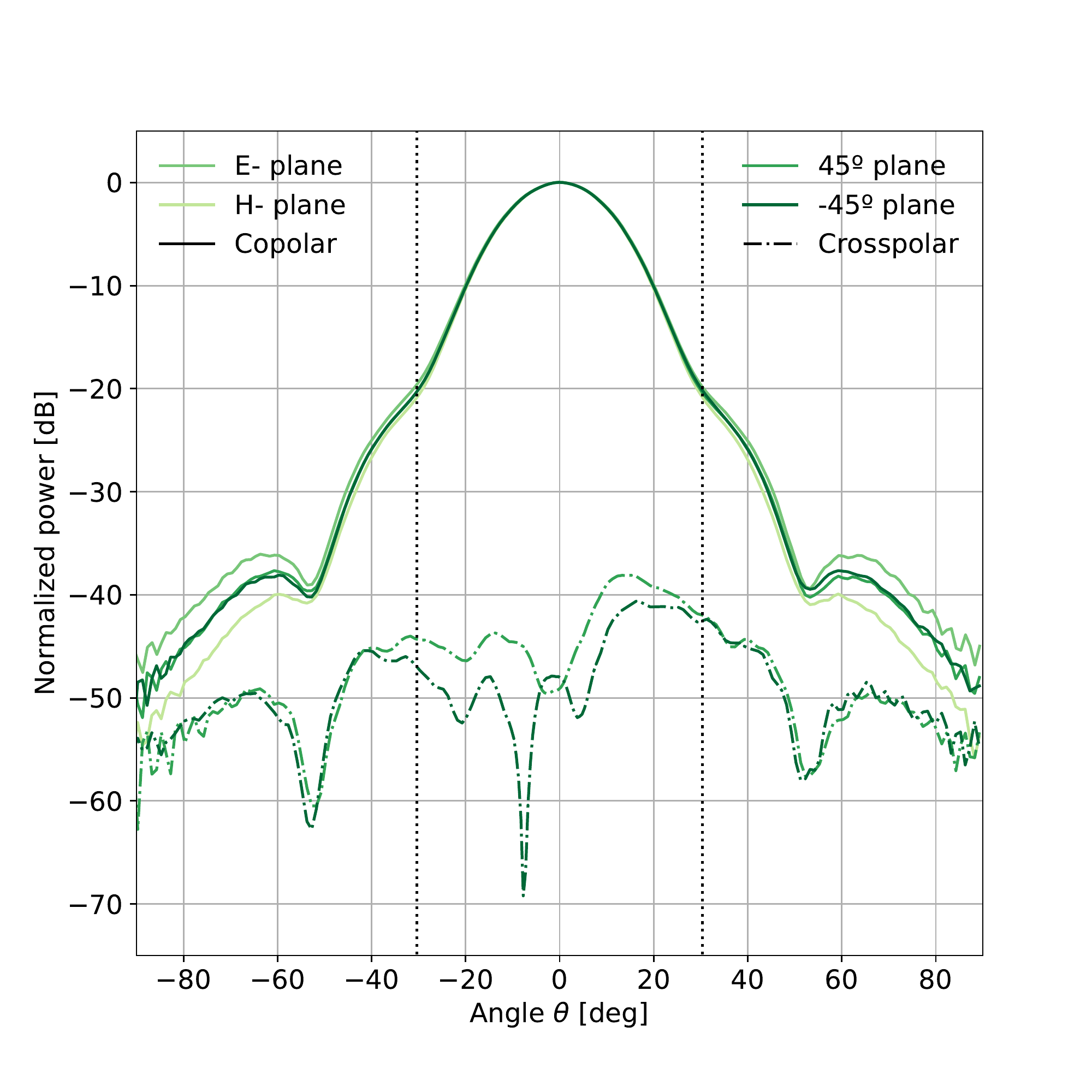}
\includegraphics[width=0.45\textwidth]{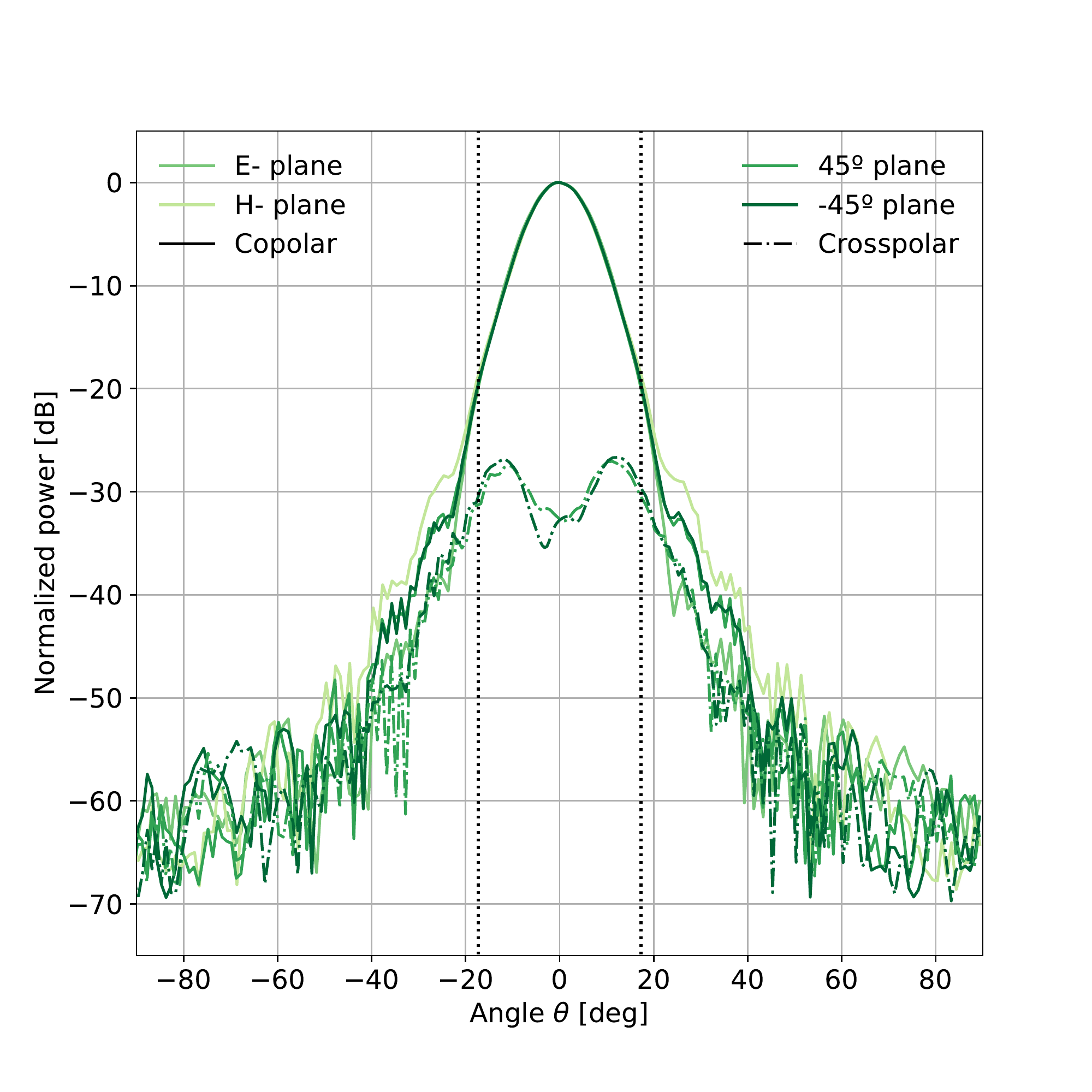}
\caption{Radiation patterns measured at the limits of the frequency band, at 10\,GHz (\emph{left}) and 20\,GHz (\emph{right}).  Copolar and cross-polar components are represented in solid and dashed lines, respectively, and in different shades for each measured plane. Vertical grey dotted lines mark the $-$20\,dB beamwidth, which varies from $\sim$60\textdegree~to 35\textdegree~across the TMS band. These figures were adapted from the work published in \cite{demiguel2021metamaterial}}%
\label{fig:meas-feedhorn-pattern}%
%\vspace{-\baselineskip}% remove one line of space below this figure caption
\end{figure*}

The antenna radiation patterns have been measured in the anechoic chamber at the University of Milano, \cite{demiguel2021metamaterial}. Using two different circular-to-rectangular waveguide transitions, i.e. WR75 and WR51, to cover the full bandwidth, they characterized each feedhorn at eleven frequencies within 10--20\,GHz. Four copolar and two cross-polar planes were measured, including the copolar E- plane, H- plane and $\mathrm{\pm45}$\textdegree~plane, and the cross-polar $\mathrm{\pm45}$\textdegree~plane.  In figure~\ref{fig:meas-feedhorn-pattern}, we show the radiation patterns at minimum and maximum frequencies in the E-, H- and $\mathrm{\pm}$45\textdegree planes.

In order to provide more information and clarity, we include table~\ref{tab:feedfeatvsfreq} with a summary of the variation of the feedhorn radiation properties over the TMS band, including the realized gain (G), the beamwidth (FWHM) at $-$3\,dB, ellipticity $e\,(\%)$, side lobe levels (SLL) and cross-polarization (XPol). Realized gain differs from the definition of gain $G$ in that it accounts for the losses due to the mismatch of the antenna input impedance ($\Gamma$) as  $\mathrm{G_r=G\cdot[1-|\Gamma|^2]}$. Beam ellipticity is defined as the percentage ratio
\begin{equation} \label{eq:ellipticity}
    e\,(\%) =
    \begin{cases}
    100\times (1-\frac{FWHM_H}{FWHM_E}), & FWHM_H \leq FWHM_E \\
    100\times(1-\frac{FWHM_E}{FWHM_H}), & \rm otherwise 
    \end{cases}
\end{equation}
being $FWHM_H$ and $FWHM_E$ the $-$3\,dB beamwidths in the principal planes. Cross-polarization remains between $-$30 and $-$40\,dB in the TMS band. 

The design of the optical system shall introduce a minimal additional cross-polarization, in order to satisfy the top-level requirement as in table~\ref{tab:specs}. On the other hand, the beam size and shape of the TMS feedhorn  change over the operating band as reported in table~\ref{tab:feedfeatvsfreq}. In general, we expect that the projected beam on the sky will present a strong dependence with frequency, as it is an  inevitable effect on ultra-wide band systems. However, the effect of the TMS feedhorn beam will be partially compensated due to the aperture size of the mirrors in wavelengths, which, in contrast to the feed beam, increases with frequency.

\begin{table*}
\centering
\caption{Variation with frequency of the feedhorn radiation properties in the TMS operational bandwidth (10--20\,GHz), including realized gain, $-$3\,dB beamwidth (FWHM) and sidelobe levels in the principal planes, beam shape or ellipticity, and cross-pol  discrimination factor.}
\label{tab:feedfeatvsfreq}
\resizebox{0.75\textwidth}{!}{%
\begin{tabular}{cccccccc}
\hline
\begin{tabular}[c]{@{}c@{}}Frequency \\ (GHz)\end{tabular} & \begin{tabular}[c]{@{}c@{}}G\\ (dB)\end{tabular} & \begin{tabular}[c]{@{}c@{}}$FWHM_{E}$ \\ (\textdegree)\end{tabular} & \begin{tabular}[c]{@{}c@{}}$FWHM_{H}$\\ (\textdegree)\end{tabular} &
\begin{tabular}[c]{@{}c@{}}$e$\\ (\%)\end{tabular} &
\begin{tabular}[c]{@{}c@{}}$SLL_{E}$\\ (dB)\end{tabular} & \begin{tabular}[c]{@{}c@{}}$SLL_{H}$\\ (dB)\end{tabular} & \begin{tabular}[c]{@{}c@{}}XPD\\ (dB)\end{tabular} \\ \hline \hline
10.0 &18.62 &22.5 &22.2 &2.55 &$-$36.26 &$-$39.81 &$-$37.33 \\
10.5 &18.96 &21.8 &21.4  &0.90 &$-$34.85 &$-$39.29 &$-$38.14 \\
11.0 &19.34 &20.8 &20.5 &1.06 &$-$34.69 &$-$37.90 &$-$40.02 \\
11.5 &19.71 &19.8 &19.8 &0.36 &$-$35.47 &$-$56.48 &$-$42.92 \\
12.0 &20.06 &18.9 &19.1 &0.47 &$-$35.95 &$-$41.39 &$-$43.74 \\
12.5 &20.45 &18.1 &18.0 &2.29 &$-$37.08 &$-$37.76 &$-$37.52 \\
13.0 &20.79 &16.9 &17.4 &3.69 &$-$34.48 &$-$38.86 &$-$40.78 \\
13.5 &21.11 &16.1 &16.8 &0.81 &$-$33.46 &$-$37.29 &$-$40.28 \\
14.0 &21.33 &16.0 &16.2 &3.81 &$-$34.90 &$-$36.67 &$-$41.78 \\
14.5 &21.59 &15.6 &16.0 &9.45 &$-$35.24 &$-$34.93 &$-$33.79 \\
15.0 &21.91 &14.9 &15.8 &9.44 &$-$33.11 &$-$35.90 &$-$34.62 \\
15.5 &22.23 &14.5 &14.8 &3.71 &$-$33.38 &$-$35.51 &$-$41.91\\
16.0 &22.46 &13.8 &14.2 &3.46 &$-$32.80 &$-$34.95 &$-$42.39 \\
16.5 &22.57 &13.4 &14.0 &4.66 &$-$32.60 &$-$35.07 &$-$38.41 \\
17.0 &22.69 &13.4 &13.9 &3.09 &$-$31.17 &$-$35.02 &$-$35.66 \\
17.5 &22.92 &13.4 &13.9 &4.02 &$-$31.11 &$-$41.36 &$-$36.06 \\
18.0 &23.32 &12.7 &13.4 &6.29 &$-$30.17 &$-$41.86 &$-$33.54 \\
18.5 &23.84 &11.5 &12.8 &9.39 &$-$27.59 &$-$33.87 &$-$26.01 \\
19.0 &23.59 &11.9 &12.1 &0.98 &$-$30.27 &$-$31.51 &$-$35.74 \\
19.5 &23.50 &12.1 &12.0 &1.97 &$-$29.64 &$-$30.63 &$-$35.77 \\
20.0 &23.54 & 12.6 &12.2 &2.89 &$-$34.70 &$-$29.02 &$-$31.97 \\ \hline
\end{tabular}%
}
\end{table*}

The first step to characterize the optical design is to determine the subtended angle of the system. We can relate this design parameter to the top-level sensitivity requirement: the subtended angle has a direct impact in the level of sidelobes (SLL) and therefore on the level of spillover, i.e. on the contribution of unwanted warm sources in the environment of the telescope (e.g. the ground) to the total system temperature, which in turn increases the noise. Minimum spillover ca be achieved by under-illuminating the parabolas, among other solutions (e.g. by using absorbing materials, avoiding blockages and ensuring clearance between optics and the supporting structure). We thus have set as a requirement for the illumination distribution so that at the edges of the mirror the illumination level is lower than $-$20\,dB, on the basis of the experience gained with the QUIJOTE experiment, \cite{QUIJOTE_SPIE2012}.
We report in table~\ref{tab:20dbbeam} the beamwidth at $-$20\,dB for both E- and H- planes of the TMS feedhorn at minimum and maximum frequency. These values give an idea of the type of reflector, --- and by type we mostly refer to the $f/D$  ratio  ---, that will be needed for the optical system to meet the requirements. We can predict that we will need a reflector with approximately 30\textdegree-half subtended angle to match  the feedhorn 20\,dB beamwidth for the worst case scenario, i.e. at 10\,GHz. This subtended angle is very wide compared to the values traditionally used in other experiments, including the QUIJOTE telescopes \cite{QUIJOTE_SPIE2012}, and it has important implications for the design of the reflector system.

\begin{table}
\centering
\caption{Beamwidth at $-$20\,dB at the principal planes of the TMS feedhorn.}
\resizebox{0.35\textwidth}{!}{%
\begin{tabular}{cccc}
\hline
                 & 10\,GHz &15\,GHz &20\,GHz \\ \hline \hline
E-plane & 59.6\textdegree    & 40.3\textdegree          & 34.6\textdegree              \\ 
H-plane & 58.1\textdegree    & 44.5\textdegree          & 35.6\textdegree              \\ \hline
\end{tabular}%
}
\label{tab:20dbbeam}
\end{table}

To conclude this section, it is worth mentioning the problem of feedhorn positioning. In narrow band systems, the feed is located in such a way that its phase center coincides with the focal point of the reflector system. We can define the phase center as the point where the feedhorn spreads the electromagnetic radiation as a spherical wave, resulting in signal with equal phase at any point on a sphere of a given radius. The radiated spherical wave is transformed into a plane wave by the reflector. A stable phase center, and its correct placement at the focal point minimizes phase error losses. However, the TMS feedhorn is a wideband antenna, and thus, the phase center location varies with frequency, causing  the defocusing at some frequency bands. Only at one particular subband the phase center of the feedhorn will be placed at the exact focal point of the reflector, while it shall be displaced at other subbands, increasing phase error losses. The final position for the feedhorn,  ---perfectly focused at the center of the band,  15\,GHz---, allows to fulfill the requirements on angular resolution, ellipticity, XPD and SLL at all frequencies.

\subsection{The cryostat window}\label{sec:cryowindow}

The TMS instrument, including feedhorns, is housed inside a cryostat that will maintain it at cryogenic temperatures below 10\,K, to isolate it from the ambient environment and ensure the low system noise temperature (or high sensitivity) requirement. A critical part of the cryostat is the vacuum window, which is necessary to separate the vacuum chamber from the environment in such a way as to ensure beam entry with low penetration loss while maintaining vacuum integrity (i.e. preventing leakage). Due the boundary between two different media such as the window and the telescope surroundings, the incoming electromagnetic (EM) waves are partially reflected, resulting in an effective power loss and other undesired effects such as standing waves and interferences that degrade the radiation performance of the system. Achieving a broadband matching between the wave impedance of the window to that of free space is key to reduce the amount of reflected light and minimize the above-mentioned effects.

The TMS cryostat is already assembled and being tested at the IAC facilities, in the absence of the window design. Traditionally, the QUIJOTE instruments have used a thin mylar film to provide low-loss insertion of the EM fields into the cryostats. Mylar is a polyester film, and it presents numerous advantages as small aperture vacuum  windows, as has been extensively discussed in the literature. However, for the TMS vacuum window we pursued an implementation based on antireflective surfaces, as an alternative to the mylar film. Two factors influence this decision: first, the extremely demanding requirements for the TMS optical system in terms of RF performance and reflection losses; and second, the required thickness for the window, which could degrade the performance if implemented with mylar. Antireflection coatings (ARCs) have been proposed as adequate alternatives, consisting of a subwavelength-period pattern etched on the dielectric surface to modify the index of refraction distribution. With an adequate pattern, any desired effective refractive index can be obtained over a given bandwidth, thus minimizing surface reflections.

Regarding RF requirements, we aimed for those designs that maximized the impedance matching, i.e. that minimized the reflection losses, and simultaneously minimized the cross-polarization contribution introduced by the window. 
We set a requirement of $-$25\,dB and $-$30\,dB for return losses ($\mathrm{S_{11}}$) and cross-polarization levels, respectively. Additionally, the cryostat window should maintain beam properties, specifically shape (ellipticity), width (FWHM) and Side Lobe Levels (SLL).

\subsubsection{Design and verification methodology}

We performed a comparative study based on the material mechanical, thermal and electrical properties, the RF performance simulated with a Finite Element Method and laboratory tests of dummy samples. 

The selection of the material took into account their permeability, rigidity and moldability, but also their electrical characteristics such as electrical permittivity and loss tangent. Ultra High Molecular Weight Poliethylene (UHMWPE) was selected to replace mylar. UHMWPE is an extremely tough and durable plastic, with excellent resistance and constant electrical characteristics over a very wide frequency range, such as low permittivity $\mathrm{\epsilon\approx 2.3}$ and low dissipation factor $\mathrm{\tan\delta\approx 1\times 10^{-4}}$  at approx. 10\,GHz,  \cite{riddle2003}. Therefore, it has been widely used in astronomy applications in the millimetric and sub-millimetric range with positive results, e.g. in the BICEP Array  \cite{hui2018}, or in ALMA receivers, \cite{schroder2015}.

\begin{figure}
    \centering
    \subfloat{\includegraphics[width=0.25\textwidth]{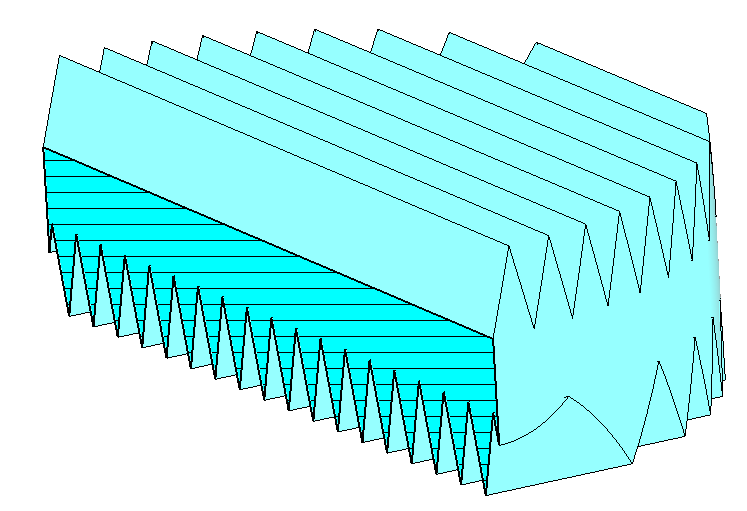}\label{subfig:pyrbed}}
    \qquad
    \subfloat{\includegraphics[width=0.15\textwidth]{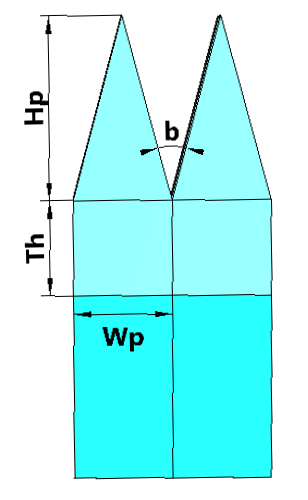}\label{subfig:pyrbed}}
    \caption{Illustration of the structure evaluated for the TMS cryostat window. \emph{Left:} Perpendicular triangular groove (PTG). \emph{Right:} Detailed view of the antireflection coatings consisting of ideal triangular grooves.}
    \label{fig:geometries}
\end{figure}

We studied the adequacy of different geometrical profiles, based on the systematic review of different structures with a special emphasis on cross-polarization presented in \cite{tapia18}. \cite{tapia18} concluded that better performance is achieved when using perpendicular geometries versus concentric structures, although the manufacturing process is more complicated. The choice of the geometry was driven first by the achieved performance and whether we met requirements, and second, by the simplicity of construction. We finally selected a pattern based on perpendicular triangular grooves. In figure~\ref{fig:geometries}, we depict the simulated antireflective layers, as well as their design geometrical parameters (thickness $Th$, wedge height $H_p$ and width $W_p$ and slant angle $b$. %The selection between each option was driven first by the achievable performance and whether we could  meet requirements, and second, by the simplicity of construction. 

We optimized the geometrical parameters of both structures to minimize reflection losses and cross-polarization over an octave bandwidth between 10--20\,GHz using CST Studio Suite. We used a genetic algorithm to optimize the selected parameters and ensure global minimums. Thickness $Th$ is a parameter driven by mechanical requirements, and it increases with the diameter of the window of the cryostat. The TMS vacuum window has a diameter of 176\,mm, so to ensure an adequate degree of deflection \textbf{ ($\mathrm{\leq 1\,mm}$)}, we needed $Th\geq$2\,mm. We designed the structures starting with a slant angle of $\mathrm{b=}$\,30\textdegree, and setting initial values for wedge height around a wavelength, at the highest frequency (20\,GHz in the case of TMS, i.e. $\mathrm{\lambda=15\,mm}$). After the optimization process, Time Domain Analysis simulations were performed to obtain the effect of the cryostat window on the feedhorn radiation pattern.  Table~\ref{tab:comparisonprototypes} gathers the results of these simulations in terms of return losses and cross-polarization. The use of Perpendicular Triangular Grooves as ARCs  means a considerable improvement in the final performance,  meeting the technical requirements by a wide margin. Return losses in the lower band do not meet the $-$25\,dB requirement, although it is mainly due to the contribution of the TMS feedhorn which. as it can be seen in the table, also fails at 10\,GHz. CST simulations of the window in isolation showed return loss values below $-$25\,dB for the whole band.

\begin{table}
\centering
\caption{Simulated reflection losses and cross-polarization for the optimized antireflection patterns in comparison with a $\mathrm{200\,\mu m}$ mylar film at 10, 15 and 20\,GHz. In the first row, we included return losses and cross-polarization levels of the TMS feedhorn for comparison.}
\label{tab:comparisonprototypes}
\resizebox{0.80\textwidth}{!}{%
\begin{tabular}{ccccccc}
\hline 
\multicolumn{1}{l}{} & \multicolumn{2}{c}{\textbf{10 GHz}} & \multicolumn{2}{c}{\textbf{15 GHz}} & \multicolumn{2}{c}{\textbf{20 GHz}} \\ \hline 
 \textbf{Subsystem} & \textbf{$S_{11}$ (dB)} & \textbf{ XPol (dB)} & \textbf{$S_{11}$ (dB)} & \textbf{XPol (dB)} & \textbf{$S_{11}$ (dB)} & \textbf{Xpol  (dB)} \\ \hline \hline
Feedhorn w/o window &$-$19.34 &$-$37.33 &$-$36.19 &$-$34.61 &$-$37.42 &$-$31.97  \\
Mylar window $\mathrm{200\,\mu m}$ &$-$18.53  &$-$35.51  &$-$24.66  &$-$35.61  &$-$20.31  &$-$28.27  \\
HDPE window w/PTG &$-$18.53  &$-$35.70  &$-$28.94  &$-$39.70  &$-$35.81  &$-$31.45 \\\hline
\end{tabular}%
}
\end{table}

In order to ensure that the layer thickness was correctly selected, the mechanical behaviour of the window was simulated using a FEM software.  Maximum directional deflection, which happens in its centre, is 0.59\,mm, below the limit of 1\,mm of of separation between the window and the IR filter (which is discussed in the section that follows) that is located directly under it. We, thus, expect no contact between both optical components.

We manufactured a window dummy with the geometrical parameters obtained in the optimization process, in order to corroborate simulation results. The IAC mechanical workshop has the capacity to manufacture these windows, using milling machines. The dummy was manufactured with PTFE (Teflon\texttrademark), as in the TMS frequency range its properties are very similar to those of UHMWPE, which was not available at that time. It presents low permittivity $\mathrm{\epsilon\sim 2.1}$ and low dissipation factor $\mathrm{\tan\,\delta\approx 3\times 10^{-4}}$ at 10\,GHz, \cite{nist1520}. Although its inferior mechanical properties presumably could make it difficult to manufacture the anti-reflective patterns, its similarity to UHMWPE helped us to make a fair comparison between the designed window and the traditional mylar film.

\begin{figure}
    \centering
    \includegraphics[width=0.5\textwidth]{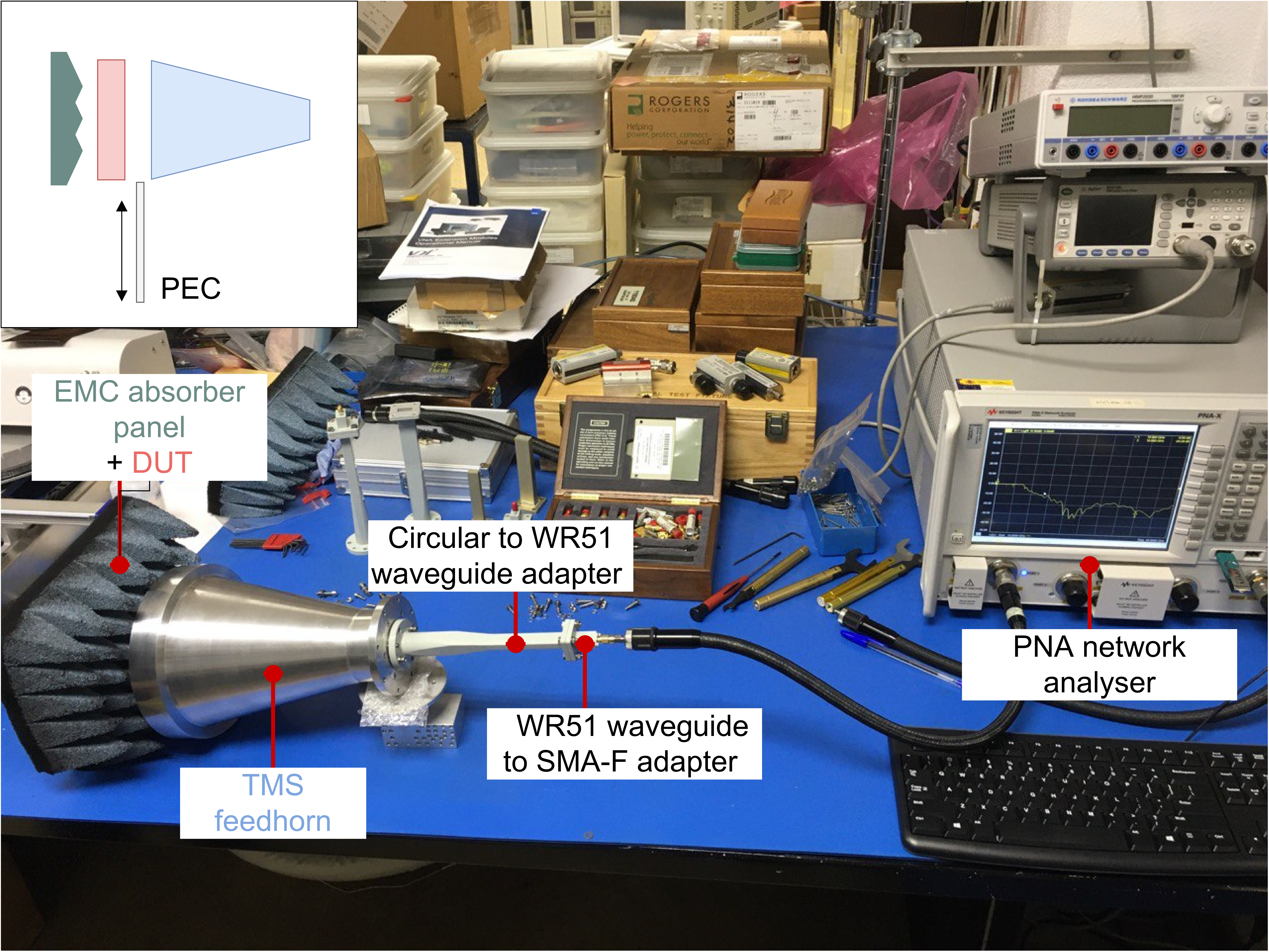}
    \caption{Setup for return loss ($\mathrm{S_{11}}$) measurements in the range of the TMS from 10 to 20\,GHz. S-parameter measurements have been taken including the feedhorn and the window, with a EMC panel installed behind it. A time gate was applied to de-embed the adaptors response. A metallic plate has been used as perfect conductor (PEC) in order to calibrate measurements.}
    \label{fig:TMSwinset}
\end{figure}

At the IAC facilities, currently we do not have the capacity to measure radiation patterns, and thus, a simple setup was conceived and assembled to obtain a measurement of return losses. Cross-polarization measurements remain pending as a future task. Measurements have been carried out for two frequency ranges, between 10--15\,GHz and 15--20\,GHz, using a PNA network analyser, the TMS feedhorn, circular (20\,mm) to WR75 and WR51 waveguide adapters, WR75 and WR51 waveguide to SMA-F adapters and, a small EMC absorber panel. We de-embedded the effect of the adaptors and isolated the measurement of the return loss $\mathrm{S_{11}}$ of the feedhorn-window subsystem by using a time domain gating technique. The setup is shown in figure~\ref{fig:TMSwinset}.

\begin{figure}
    \centering
    \includegraphics[width=0.8\textwidth]{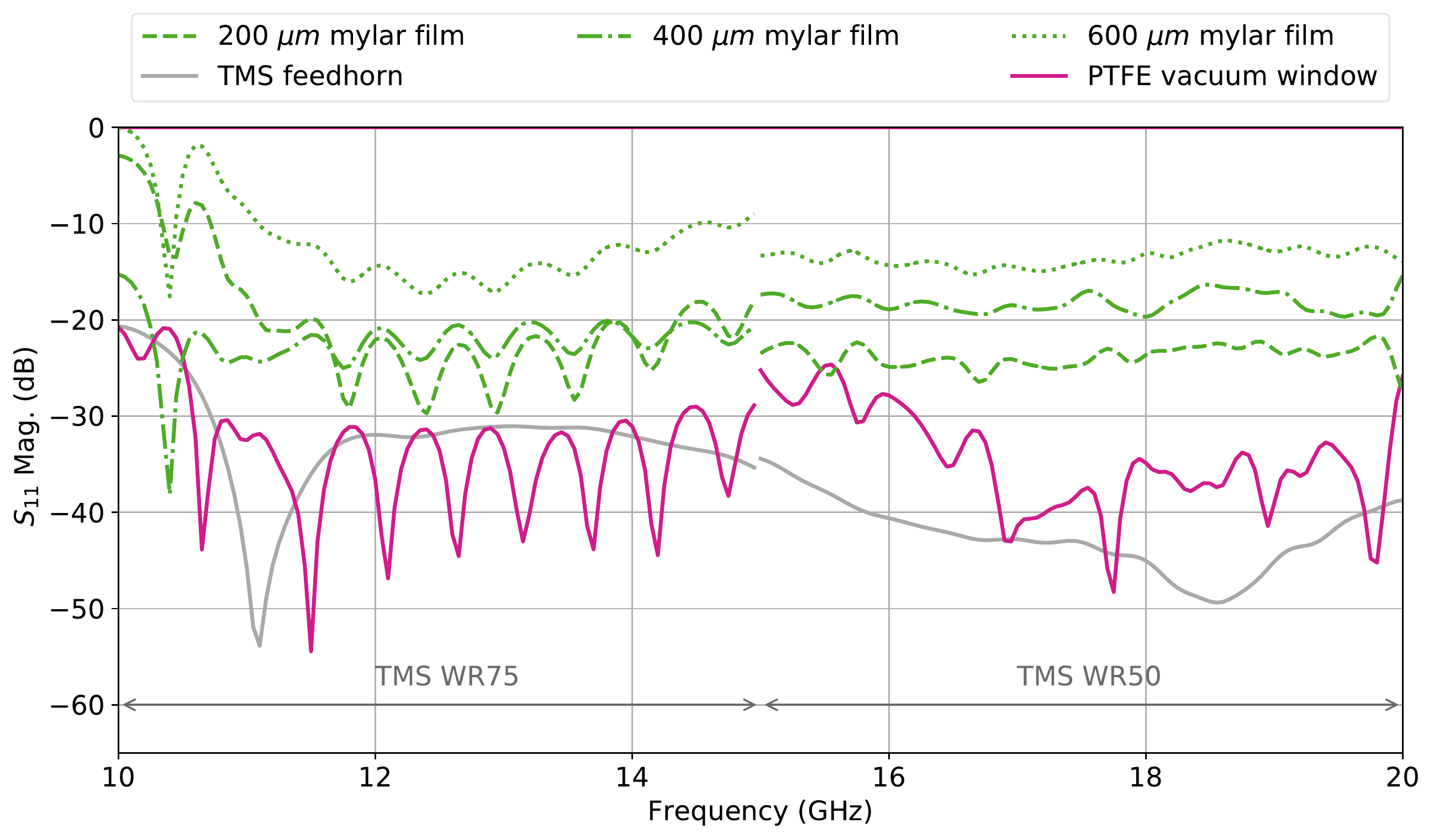}
    \caption{Measurement of the return loss (dB) of the prototype window (\emph{pink solid line}) and mylar films  (\emph{green non-solid lines}) with different thicknesses in the TMS frequency range. Return loss of the stand-alone TMS feedhorn  has been included to highlight the effects of the window. Two different setups have been used to fully cover the TMS range, measuring between 10--15\,GHz (WR75) and 15--20\,GHz (WR51).}
    \label{fig:winmeascomparison}
\end{figure}

Measurements, plotted in figure~\ref{fig:winmeascomparison}, show a substantial improvement in the behaviour of PTFE windows with anti-reflection patterns with respect to mylar films. This improvement is more noticeable going up in frequency, where mylar films present return losses over $-$20\,dB. By contrast, PTFE windows present return losses below $-$30\,dB along the whole band, only peaking at the upper and lower limits of the frequency ranges, which may be caused by the time gating function. In the case of the TMS, different values for the $\mathrm{S_{11}}$ at central frequency 15\,GHz could have been introduced by the different setups (WR75 or WR51 waveguide) as well as the time gating operation.

\subsubsection{Vacuum window final design}

The TMS vacuum window is a panel of Ultra-High-Molecular-Weight Polyethylene (UHMWPE) with a diameter of 176\,mm and 2\,mm thickness. The surface of both sides have a pattern of triangular grooves in perpendicular directions. The dimensions of these grooves, as optimized with CST Studio Suite, were 10.4\,mm and 19.4\,mm for the height and width, respectively. In order to guarantee sufficient stiffness of the window, its outer perimeter thickness is kept equal to the triangle height on the air side.

A full picture of the window RF performance and its effect on the radiation pattern, including transmission losses and beam characterization (cross-polarization levels, beam shape and size, etc.), was not possible at the moment with the available resources at the IAC facilities. However, both $\mathrm{S_{11}}$ measurements and CST simulation results corroborate the superior behaviour of PTFE/UHMWPE windows over mylar films, and thus, it is reasonable to think that they are adequate for the scientific scope of TMS.

\subsection{IR filter}\label{sec:feedext}

\begin{figure}
    \centering
    \includegraphics[width=0.4\textwidth]{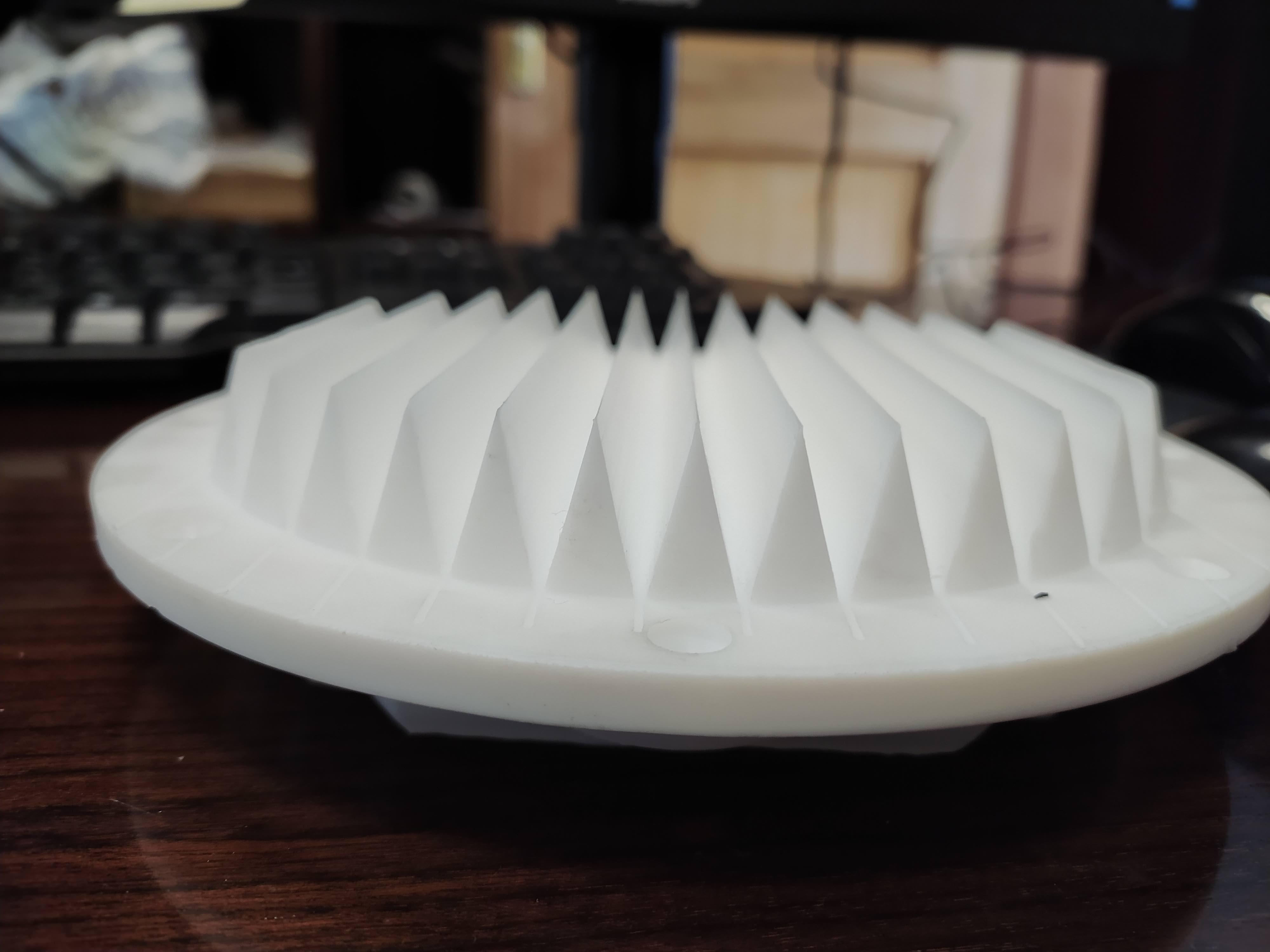}
    \caption{Photograph of the TMS IR filter. The dimensions of the groove pattern were optimized with CST Studio and subsequently machined at the IAC mechanical workshop using milling cutter machining. Total thickness $Th$ is 2\,mm.}
    \label{fig:filter}
\end{figure}

The Infrared Radiation (IR) filter has been implemented using Polytetrafluoroethylene (PTFE or Teflon\texttrademark) and a similar design as the vacuum window, as shown in figure~\ref{fig:filter}. The same triangular groove patterns have been machined in the filter in order to increase the throughput of the sky signal in the TMS band, and reduce reflections. In addition, while the vacuum window will be at room temperature, as it is the medium separating the inside of the cryostat from the outside, the IR filter will be at 10\,K.  This means that, although the dissipation factor of PTFE is approximately twice that of HDPE, as the filter is at cryogenic temperatures its resulting transmission losses will be reduced, \cite{jacob2002}, and  we can assume them to be at comparable levels.

Regarding IR frequencies absorbance, PTFE has been widely used to implement IR filters in ALMA receivers as it provides good absorbance levels without requiring large thicknesses. Infrared transmittance and emittance of PTFE has been widely characterised taking sheet thickness and operating temperature into account by \cite{jacob2002}, \cite{Wentink1961}, and \cite{liang1956}.

\subsection{The TMS dual-reflector system} \label{sec:reflectorconfiguration}

To this day, CMB telescopes have generally used decentered (also called ''off-axis'' or ''offset-fed'') reflector-based optical systems  \cite{hanany2013}. These off-axis systems, although less compact, present a major advantage over the centered version, as there is no natural self obscuration by the reflectors or the feedhorn. As a result, the radiation scattering is reduced, a great advantage for CMB absolute measurements. Therefore, we evaluated  several decentered optical configurations until the final design of the TMS optical system was found. In particular, we  studied the adequacy of a combination of the TMS feedhorn and a single reflector, or a dual-reflector system in two different configurations, Crossed-Dragone and Gregorian. 

Most designers agree that, although compactness is a valuable feature, a single reflector is often too limited and thus, many designs rely on two or three reflectors. Regardless of this fact, we started considering the single reflector option, seeking to simplify and make the design more compact, and even reuse the reflector from the old VSA experiment \citep{VSA1}. The radiation properties of the feedhorn (see section~\ref{sec:TMSfeed}) and the physical limit imposed by the cryostat to avoid obscuration completely discarded the possibility of using the VSA reflector. It also led to a design based on a 1.5\,m parabolic reflector with an offset angle of 60\textdegree. This configuration provided very adequate sidelobes and cross-polarization levels, but the radiated beam resulted overly elliptical. To achieve polarimetric capability, as was mentioned  in the previous section, we favour highly symmetrical beams, and therefore, the single reflector configuration was discarded.

We evaluated the most common dual-reflector configurations. The mainstream two-reflector configurations are Cassegrain (parabolic and hyperbolic mirrors) or Gregorian (parabolic and elliptic mirrors).  \cite{dragone1978} provides a useful method for designing reflector antennas with low cross-polarization levels. Moreover, the Cross-Dragone configuration has been widely used in CMB and polarimetric experiments, including the QUIJOTE telescopes, \cite{QUIJOTE_SPIE2012}, with successful results. However, the Cross-Dragone system is optimized for feed beamwidths of less than 25\textdegree. The TMS feedhorn presents a larger beamwidth to ensure illumination levels of $-$20\,dB (see section~\ref{sec:TMSfeed}), preventing us from using this configuration. In practice, this feature translated into the need for a more compact mirror system, with a much smaller focal ratio than that of the QUIJOTE system.

We therefore opted for an optical system similar to the Planck satellite, \cite{tauber2010}, using a Gregorian paraboloid-ellipsoid configuration. The design originally meets the Dragone-Mizuguchi condition of minimum cross-polarization, \cite{dragone1978,mizugutch1976}. The design was finely tuned to ensure this  to the maximum extent possible, while taking into account the RF properties of the feed and the physical dimensions of the cryostat, which was already fabricated. The work methodology included a previous analytical study with the Student Edition of GRASP\footnote{\url{https://www.ticra.com/software/grasp-student-edition/}} to ensure the Mizuguchi condition. This simulation was straightforward and did not take into account the RF properties of the TMS feed or typical effects on the radiated beam due to diffraction. Moreover, it did not take into account the presence of the cryostat, which acts as an additional diffractor and can cause blockage. Therefore, the next step included the verification of the 3D mechanical model and subsequent adjustments.  This process required a few iterations until the final design was obtained. To conclude, we performed a more complete analysis of the system radiation properties acknowledging the aforementioned diffraction effects and the TMS feed behaviour. To this end, we used  CST Studio Suite and performed a preliminary study assuming an ideal system, which is presented in section~\ref{sec:preliminarstudy}.

\begin{figure}
    \centering
    \includegraphics[width=0.75\textwidth]{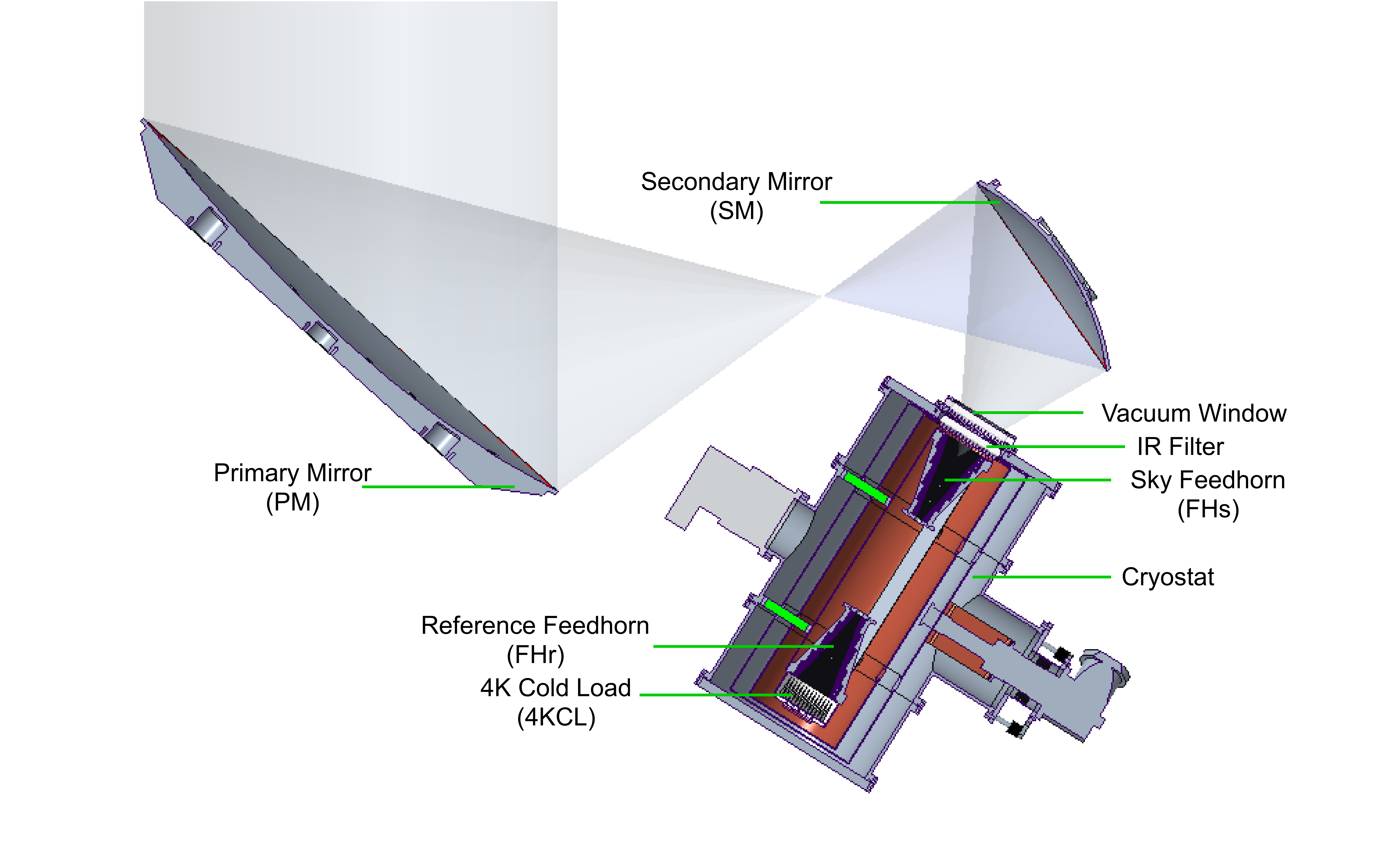}
    \caption{3D visualization of the TMS optical system, as  modelled with the Creo Parametric 3D Modelling software. The optical design is based on a Gregorian offset configuration, with a 1.49\,m-primary mirror and a 0.6\,m-ellipsoidal secondary mirror. The TMS feedhorn is positioned at one of the sub-reflector foci, aligned with its phase center at 15\,GHz. The removal of this sub-reflector allows us to switch between observations with and without optics. }
    \label{fig:optconf}
\end{figure}

Figure~\ref{fig:optconf} shows the final optical design, based on a Gregorian system with a projected diameter aperture of 1.2\,m, and  an equivalent focal length of 0.7\,m, resulting in an F/D\textasciitilde 0.58. The primary mirror (PM) of the TMS is an offset paraboloid, with 1.49\,m major axis, and the secondary mirror (SM) is an offset concave ellipsoid, with a 0.6\,m diameter and eccentricity $\epsilon$ of 0.45. The half-angle subtended by the SM at the feed is 30.32\textdegree, ensuring illumination levels $\leq-$20\,dB in the mirror rims along the whole TMS band. The system provides an angular resolution of about 1.5\textdegree\,on average.

It is worth noting that the final system does not satisfy the Mizuguchi \cite{mizugutch1976} condition of minimum cross-polarization by $\sim$15\textdegree~of deviation with respect to the optimal pointing of the antenna.  Similarly, the condition given by Rusch \cite{rusch1990} for minimum cross-polarization and spillover is not met. This design decision was motivated by the need to separate the cryostat and its pump as far as possible from the optics, in order not to produce a blockage. Shading by the cryostat was avoided by slightly reducing the pointing angle of the antenna with respect to the PM axis. Nevertheless, the cross-polarization introduced by the optical system remains very low, barely degraded with respect to the optimal configuration, according to GRASP results. In the same way, spillover losses are almost negligible in comparison to the optimal configuration. Table~\ref{tab:mizuguchi} presents the design parameters of the optical system, including the optimal and chosen values for the PM offset angle and the angle between the feed and SM axis. Optimal values are extracted from the named conditions, respectively given by the expressions
\begin{equation}
\label{eq:miz}
\tan \alpha = \frac{|\epsilon^2-1|\sin \beta}{(1+\epsilon^2)\cos \beta-2\epsilon}
\end{equation}
and
\begin{equation}
\label{eq:rus}
\tan \frac{\beta}{2}=\left (\frac{\epsilon-1}{\epsilon+1}\right )^2 \tan \left ( \frac{\beta+\theta_0}{2}\right )
\end{equation}
where $\mathrm{\alpha}$ is the angle between the feed pointing direction and the SM axis of symmetry; $\mathrm{\beta}$ is the angle between the SM and PM axes; $\epsilon$ is the SM eccentricity, defined as the ratio of the minor axis to the major axis; and $\theta_0$ is the offset angle of the PM, which coincides with the pointing angle of the feed antenna after the first reflection on the secondary mirror.

\begin{table}
\centering
\caption{Mirror design parameters for the TMS optical system. Optimal values for the angle between the sub-reflector and primary mirror $\mathrm{\alpha^{opt}}$ and the PM offset angle $\mathrm{\theta_0^{opt}}$ were obtained from the Mizuguchi and Rusch formulations.  The differences between optimal and final values were introduced to avoid blockage of the cryostat, at the expense of slightly degrading XPol and spillover characteristics.}
\label{tab:mizuguchi}
\resizebox{0.7\textwidth}{!}{%
\begin{tabular}{ccc}
\hline 
\multicolumn{1}{c}{Parameter description}        & \multicolumn{1}{c}{Symbol}   & \multicolumn{1}{c}{Value} \\ \hline \hline
\multicolumn{1}{c}{Eccentricity}                 & \multicolumn{1}{c}{$\epsilon$}        & \multicolumn{1}{c}{0.45}  \\ 
\multicolumn{1}{c}{Angle between PM and SM axis} & \multicolumn{1}{c}{$\mathrm{\beta}$}     & \multicolumn{1}{c}{40\textdegree}    \\ 
\multicolumn{1}{c}{Optimal angle between SM and feed axis (from eq.~(\ref{eq:miz}))} & \multicolumn{1}{c}{$\mathrm{\alpha^{opt}}$} & \multicolumn{1}{c}{$-$87.63\textdegree} \\ 
\multicolumn{1}{c}{Optimal PM offset angle (from eq.~(\ref{eq:rus})) }              & \multicolumn{1}{c}{$\mathrm{\theta_0^{opt}}$} & \multicolumn{1}{c}{96.86\textdegree}   \\ \hline
\multicolumn{1}{c}{Angle between SM and feed axis}         & \multicolumn{1}{c}{$\mathrm{\alpha}$}      & \multicolumn{1}{c}{$-$72\textdegree}   \\
\multicolumn{1}{c}{PM offset angle}              & \multicolumn{1}{c}{$\mathrm{\theta_0}$} & \multicolumn{1}{c}{70.5\textdegree}   \\
\hline     
\end{tabular}%
}
\end{table}

\begin{figure}
    \centering
    \includegraphics[width=0.75\textwidth,trim={0 2cm 0 1cm}]{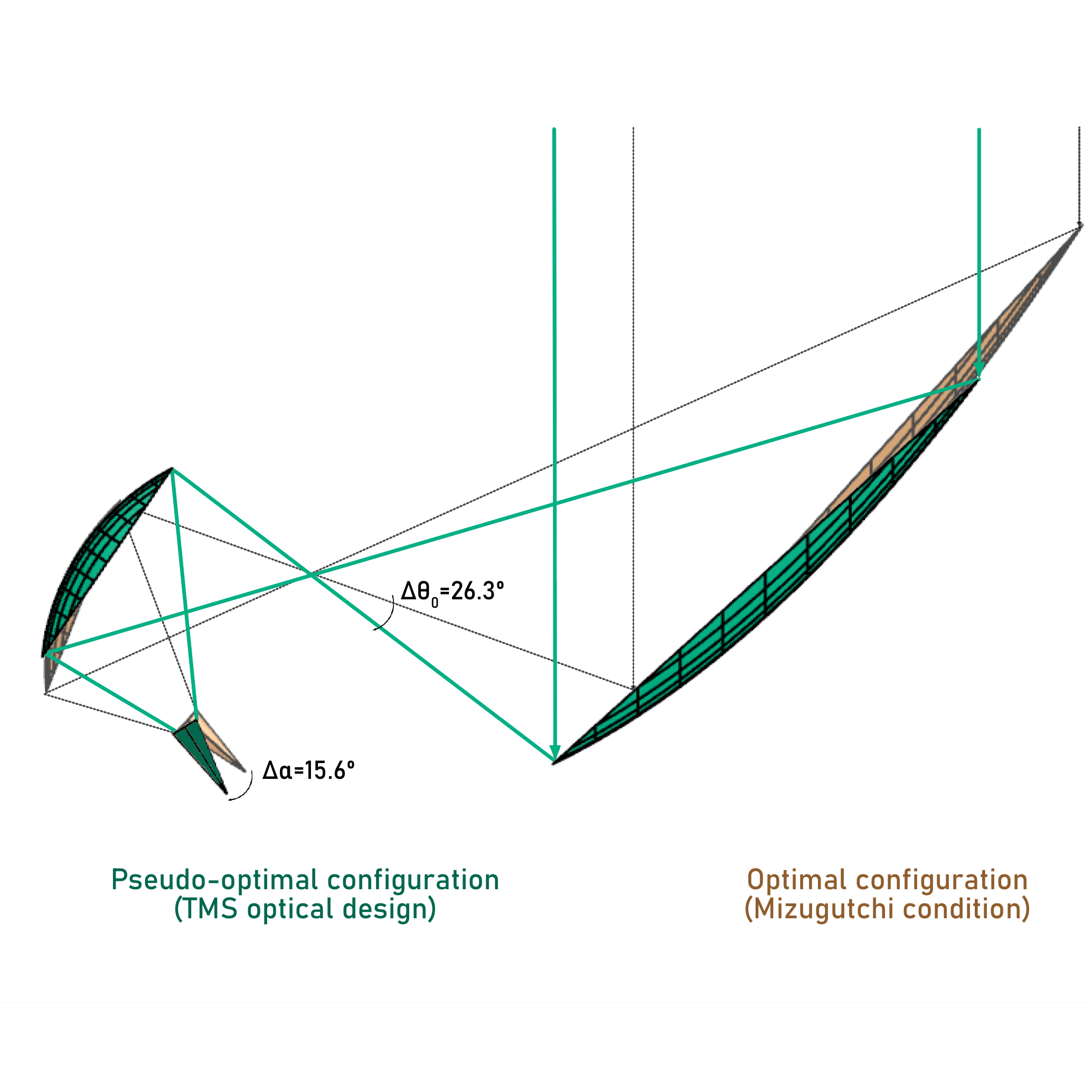}
    \caption{ Illustration of the initial or optimal configuration (\emph{orange colour}, fulfilling the Mizuguchi condition) and the near-optimal configuration (\emph{green colour}) achieved for the TMS optical system after  appropriate iterations. In this scheme, only the differences between the two configurations are highlighted, while table~\ref{tab:mizuguchi} shows the values for each design parameter.}
    \label{fig:mizuguchi}
\end{figure}

%%%%%%%%%%%%%%%%%%%%%%%%%%%%%%%

\section{The TMS optical performance}\label{sec:preliminarstudy}

Here we describe the performance of the optical system, derived from CST Studio Suite simulations. We have used a 3D full-wave solver, the Integral Equation Solver (IES), which presents great efficiency when simulating large models, as in this case. All simulations shown in this section, unless otherwise stated, have been made including the TMS feedhorn and the dual-reflector system, but the  3D model does not include the support structures or the vacuum window, to reduce unnecessary computational complexity. However, we do include a simplified version of the cryostat to take into account the possible adverse effects of the vacuum pump, shown in figure~\ref{fig:optconf}, which is the structure closest to the field of view of the mirrors. The feedhorn was excited with a single mode with linear 0\textdegree and 90\textdegree~polarizations. 

This study will complement the beam measurements during the comissioning phase,  providing valuable information on systematic errors for the scientific exploitation phase. We start with an evaluation of the level of sub-illumination of the mirrors, in direct relation to the illumination and spillover efficiencies, and thus, to the shape and size of the radiation pattern. We continue with the in-depth study of the main beam and far sidelobes, with special emphasis on frequency variation. The TMS instrument is a spectrometer that will continuously acquire the sky spectrum and compare it to the spectrum of a cold calibrator \cite{rubino2020, alonsoarias2020}. While the spectrum of the calibrator is known and stable over the TMS frequency range, it is necessary to have perfectly defined and characterised the frequency response of the instrument, including the optical system.  We conclude this study with an analysis and estimation of the polarization leakage when obtaining the Stokes parameters spectra.

\subsection{Evaluation of  illumination levels on the mirror edges }\label{sec:edge}

A correct illumination level is key to ensure performance in terms of spillover, i.e. to ensure the mitigation of undesired contributions that could contaminate the absolute measurement of the sky temperature.  In radioastronomy and remote sensing applications, sidelobes reduction is ensured firstly by sub-illuminating the mirrors, and secondly, by using shielding structures. Following the selection of the optical configuration, our first step was to evaluate the illumination levels on the mirror rim. 

\begin{figure}
    \centering
    \includegraphics[width=0.65\textwidth]{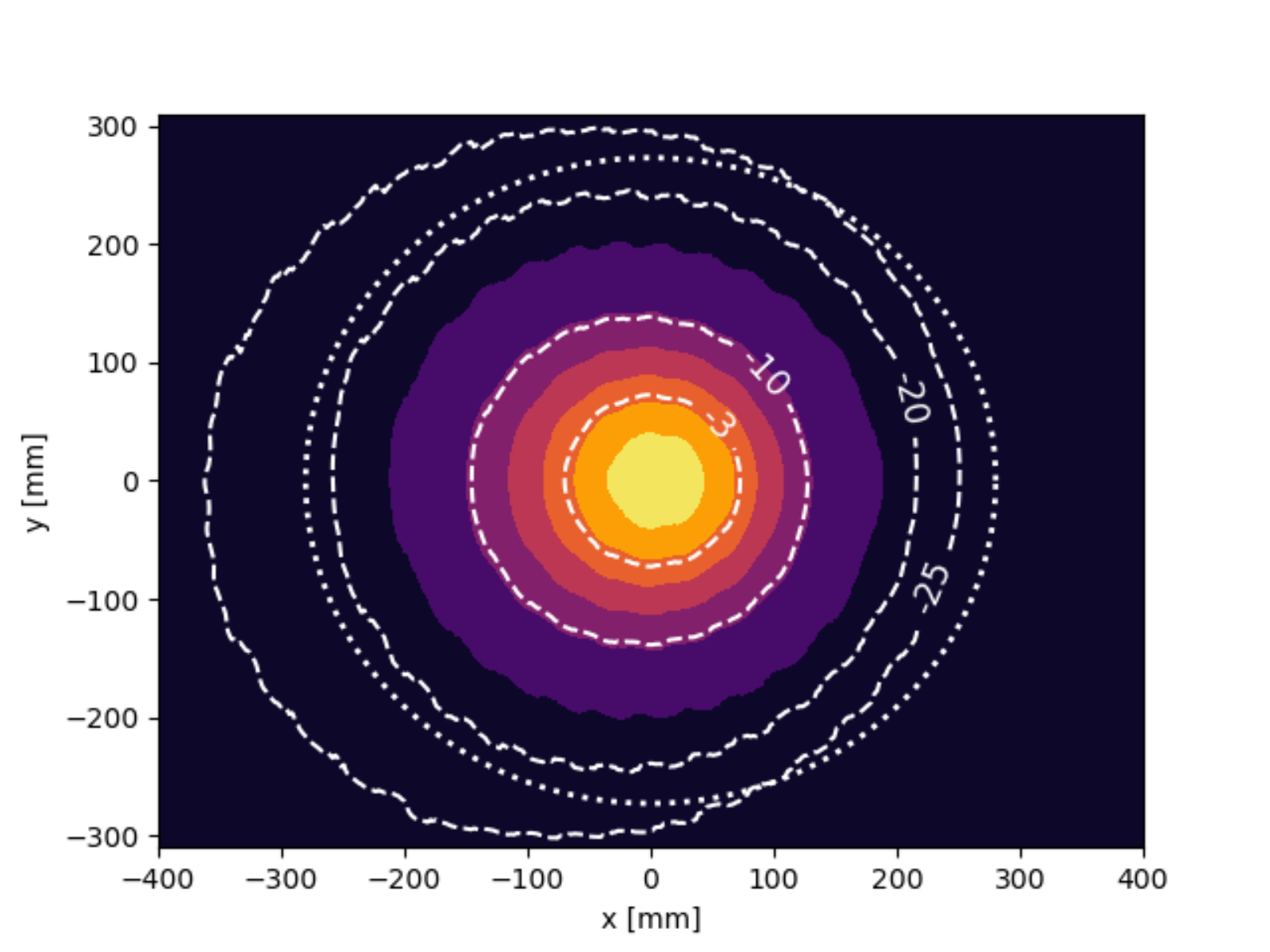}
    \caption{Near field (NF) contour lines in the plane that gathers  the secondary mirror edges at 10\,GHz. The dotted line represents the ellipse that forms the edge of the sub-reflector. An illumination taper of between $-$20 and $-$25\,dB is verified along the entire border of the mirror.}
    \label{fig:tapervariation}
\end{figure}

From the Near Field (NF) simulation of the TMS feedhorn, the incident field on the subreflector surface has been computed and represented. The total field amplitude is defined in terms of the field vector magnitude in the $(x,y,z)$ coordinate system, where the $z$-axis is the direction of maximum radiation. Figure~\ref{fig:tapervariation} shows the field distribution for the lower limit of the TMS operational bandwidth, i.e. 10\,GHz. In addition to the contour lines, we have represented the SM border as a white dotted line. It can be seen that the illumination level peaks in the mirror center and drops rapidly towards values between $-$20 and $-$25\,dB at the mirror edges. The illumination distribution is roughly elliptical and so, we obtained non-constant tapers at the mirror borders. On the other hand, the farfield  radiation patterns of the TMS feedhorn, included in section~\ref{sec:TMSfeed}, show a clear dependence with frequency, meaning we will observe this dependence also on the edge taper. The effects of this spectral variation are studied in the following subsection. 

\begin{figure}
    \centering
    \subfloat{\includegraphics[width=0.45\textwidth]{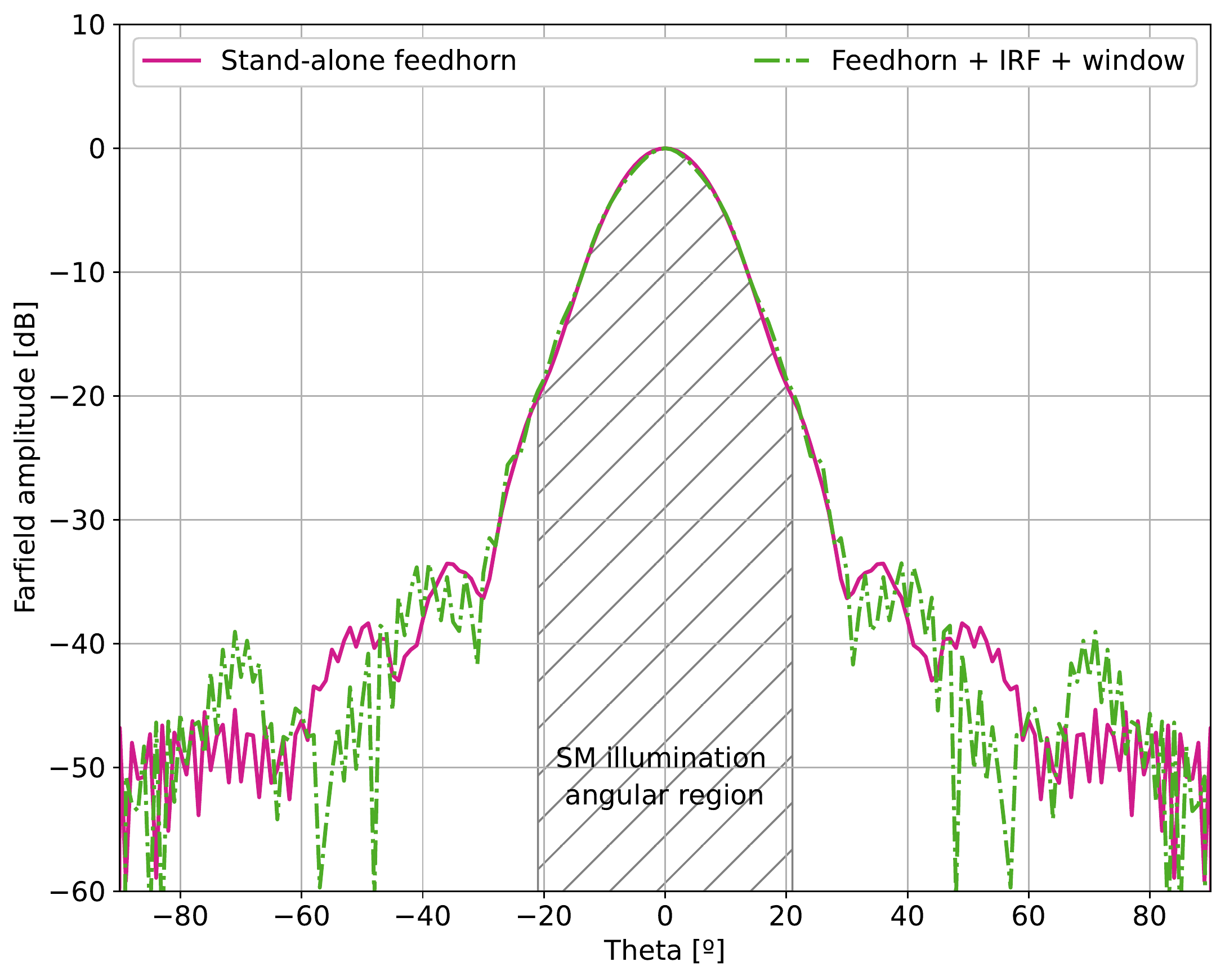}}
    \subfloat{\includegraphics[width=0.45\textwidth]{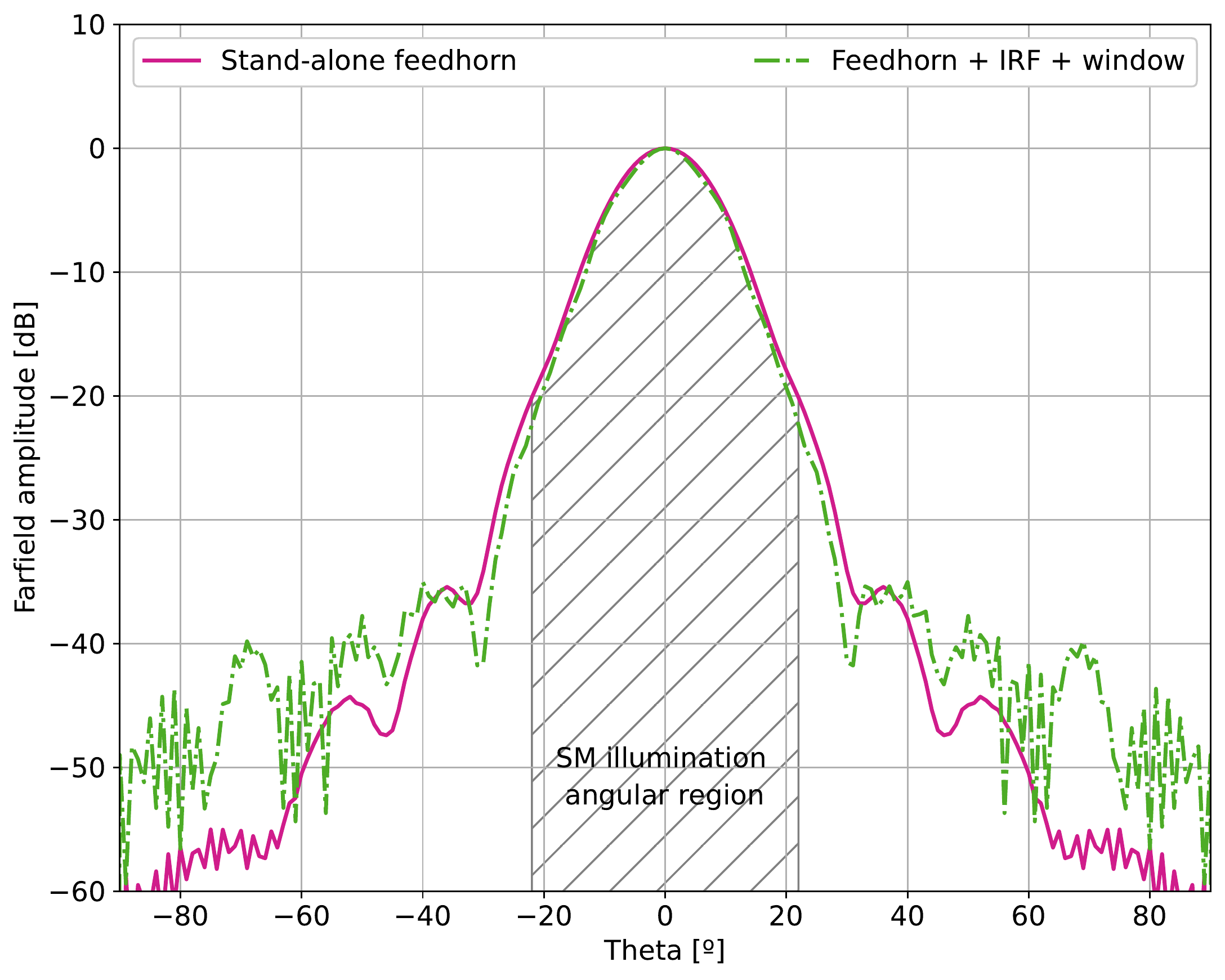}}
    \caption{ Farfield cuts at $\mathrm{\phi=0^{\circ}}$ and $\mathrm{90^{\circ}}$ in the center of the TMS band, at 15\,GHz. We compare the radiation pattern of the stand-alone TMS feedhorn (in\emph{ pink colour}) and the radiation pattern of the feedhorn with the infrared filter (IRF) and the cryostat window (in\emph{ green colour}). The \emph{hatched} angular region corresponds to the secondary mirror (SM) illumination region.}
    \label{fig:comparison_cut_feedwin}
\end{figure}

On a separate note, the TMS vacuum window and IR filter have not been considered for the computation of the NF, due to the limited computing resources. We, therefore, include table~\ref{tab:featsfeedcomp} and figure~\ref{fig:comparison_cut_feedwin} to assess their effects on the radiation pattern. Table~\ref{tab:featsfeedcomp}  gathers some figures of merit, including cross-polarization and sidelobe levels, realized gain,  beamwidth at $-$20\,dB and ellipticity, at 10, 15 and 20\,GHz. The radiation pattern principal cuts at the center of the TMS band, 15\,GHz, are shown in figure\,\ref{fig:comparison_cut_feedwin}. Radiation properties do not change significantly, keeping the beam shape and cross-polarisation level within the established margins.  Therefore, we can safely conclude that the illumination levels at the secondary mirror still comply with the $-$20\,dB requirement. The sidelobe level is affected with changes of up to 15\,dB, although these correspond to small sidelobes directly next to the main beam. However, these values are excellent to start with and it is possible that the simulation does not model them well due to the complexity of the arrangement. If these values are a worst case, then they are still acceptable in terms of antenna performance.

\begin{table}
\centering
\caption{Comparison of the effects of the vacuum window and the IR filter in the RF performance of the TMS feedhorn, in the TMS frequency range, at 10, 15 and 20\,GHz. Evaluated properties include: directivity $\mathrm{D_0}$\,(dB), cross-polarization XPD (dB), sidelobe levels SLL (dB), mean beamwidth BW at $-$20\,dB (\textdegree) and beam ellipticity e\,(\%). 
}
\label{tab:featsfeedcomp}
\resizebox{0.8\textwidth}{!}{%
\begin{tabular}{lllllll}
\hline
Frequency & System & $\mathrm{D_0}$\,[dB] & XPD\,[dB] & $\mathrm{BW_{-20\,dB}}$\,[\textdegree] & e\,[\%] & SLL\,[dB] \\ \hline\hline
\multirow{2}{*}{10\,GHz} & Feedhorn (FH) &18.68 &$-$37.33 &58.83 &1.37
&$-$38.04  \\
 & FH + IR filter + Window &18.62 &$-$33.09 &57.71 &4.13 &$-$22.38 \\ \hline
\multirow{2}{*}{15\,GHz} & FH &22.23 &$-$39.24 &42.42 &2.65 &$-$34.51  \\
 & FH + IR filter + Window &21.54 &$-$38.37 &41.89 &2.77 &$-$35.26  \\ \hline
\multirow{2}{*}{20\,GHz} & FH &23.51 &$-$31.97 &35.15 &5.19 &$-$31.86  \\
 & FH + IR filter + Window &23.59 &$-$29.49 &32.41 &4.52 &$-$21.37  \\ \hline
\end{tabular}%
}
\end{table}

\subsection{Main beam: behaviour with frequency}\label{sec:radiationfreq}

\begin{figure}
    \centering
    \includegraphics[width=0.8\textwidth]{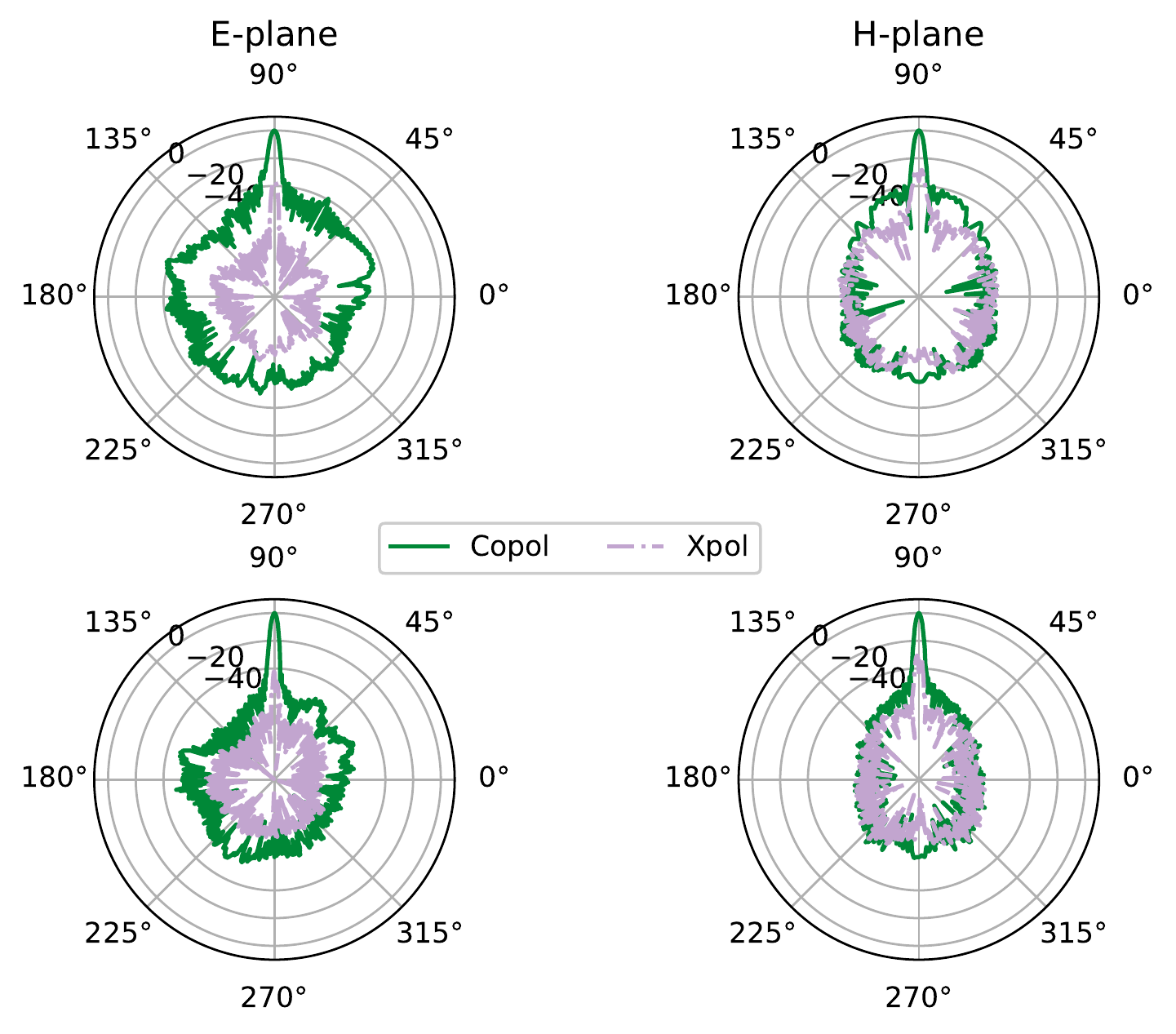}
    \caption{TMS antenna principal plane patterns at 10 (\emph{top}) and 20\,GHz (\emph{bottom}). For each cut E- (\emph{left}) and H-plane (\emph{right}), we computed the co-polar (in \emph{green colour}) and cross-polar components (in \emph{violet colour}). The direction of maximum radiation in   $(\theta,\phi)=$~(90\textdegree,90\textdegree) coincides with the zenith in the telescope's coordinate system. The radiation patterns have been normalised to the directivity value in each subband, for the sake of comparability. Tables~\ref{tab:beamvsfreq_0pol} and~\ref{tab:beamvsfreq_90pol} gather information on the directivity achieved in each band.}
    \label{fig:polarpatterns}
\end{figure}

In a wideband system such as the TMS, the expected beam shape will be strongly dependent on the frequency band. The TMS feedhorn response introduces a frequency-dependant variation on the mirror edge taper, as described in sections~\ref{sec:TMSfeed} and~\ref{sec:edge}. The resulting decrease of illumination at higher frequencies is partially counterbalanced with the variation of the wavelength-mirror diameter ratio, but we must characterise this effect in detail. We, therefore, have run beam simulations between 10--20\,GHz, in subbands of 0.5\,GHz. A brief summary of the simulation results  is presented in 
Figure~\ref{fig:polarpatterns}.  We compared both copolar and cross-polar patterns in the principal planes of the antenna for 10 and 20\,GHz. We have also fitted the copolar pattern to a 2D-Gaussian to accurately quantify the main beam FWHM, obtaining a variation between the TMS band limits of between 126.61  and 97.90\,arcmin.

\begin{figure*}
    \centering
    \includegraphics[width=0.7\textwidth]{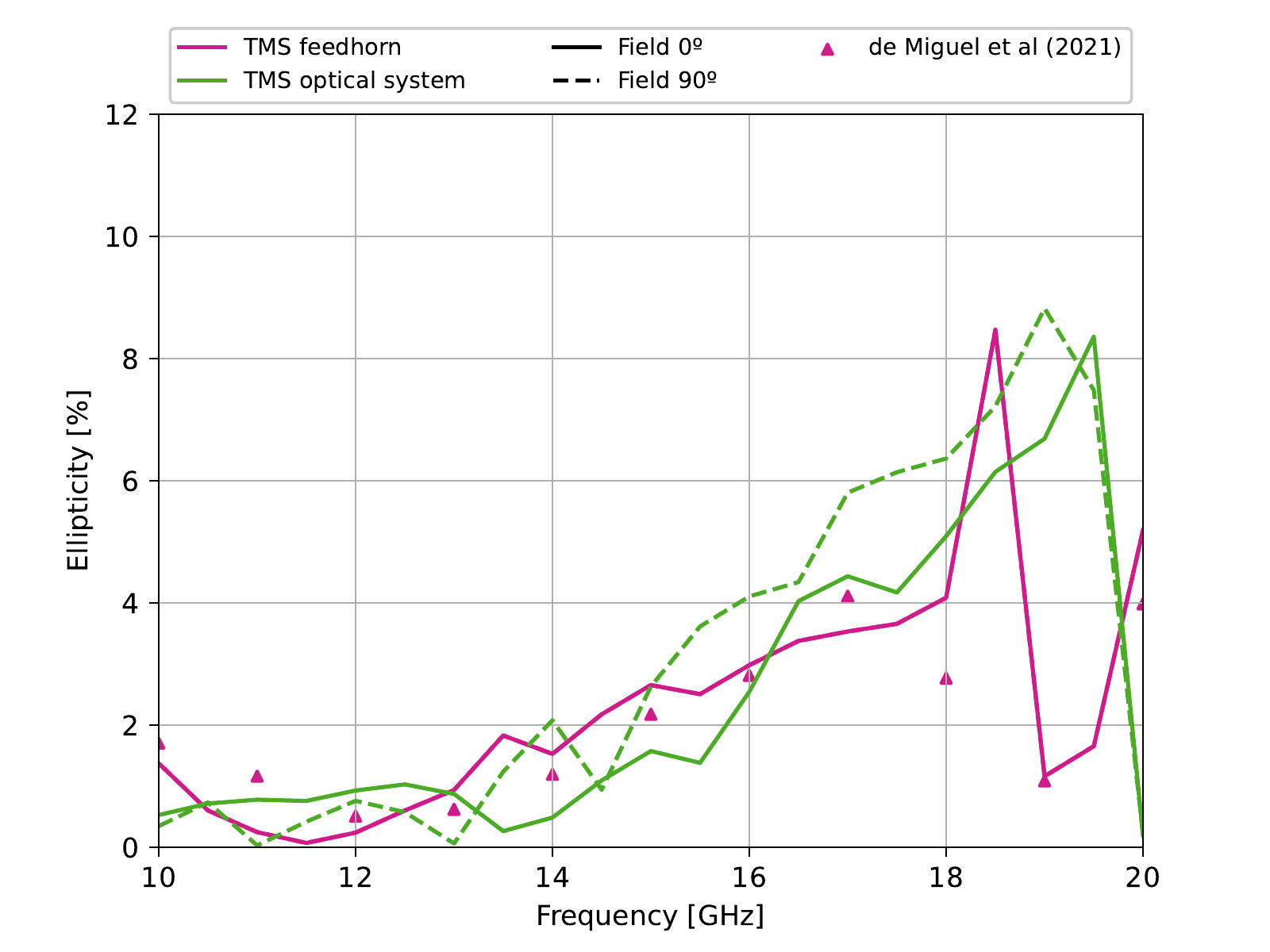}
    \caption{Variation over frequency of the beam ellipticity when operating with the complete optical system (\emph{green line}) or merely with the feedhorn (\emph{pink  line}), in the cases of 0\textdegree~(\emph{solid line}) and 90\textdegree~(\emph{dashed line}) polarizations. The scatter of \emph{triangles} mark the ellipticity values inferred from the FF measurements made by \cite{demiguel2021metamaterial}. }
    \label{fig:comparisonellipticity}
\end{figure*}

We report some relevant radiation characteristics in tables~\ref{tab:beamvsfreq_0pol} and~\ref{tab:beamvsfreq_90pol}, drawing a distinction between linear polarisation at 0 and 90\textdegree. Across the band, the beam directivity presents a total change of 10\%, which is a small improvement with respect to the 22\% variation presented by the feedhorns. The cross polar discrimination factor (XPD) is at least $-$30\,dB for the whole operating band, and does not drop below $-$35\,dB in the upper part of the frequency range ($\mathrm{\nu\geq15\,GHz}$). The ellipticity is kept under 5\% in the first half band, and we obtain the highest value (an 8\%) at 19\,GHz. The possibility of the reflector being 
the sole case of the difference between the beam elongations in the two directions was dismissed when we calculated the beam ellipticity of the TMS feedhorn, which is displayed in figure~\ref{fig:comparisonellipticity}. We show the ellipticity calculated from the CST simulations using the expression (\ref{eq:ellipticity}) for both the horn and the optical system. In addition, we used the same method to calculate the ellipticity from the reported far-field (FF) measurements in \cite{demiguel2021metamaterial}. The behaviour of the horn in bands above 18\,GHz is explained by \cite{demiguel2021metamaterial}, which attributes it to the impossibility of maintaining a single fundamental mode ($\mathrm{TE_{11}}$) in the whole 10--20\,GHz. From approx. 18\,GHz, the excitation of higher-order modes can explain the slight change in beam shape, which becomes more spherical, if we look at both the simulation and measurements results. In any case, the effect of these higher-order modes is not detrimental to the behaviour of the whole optical system\footnote{So far, we have run the simulations without considering a multi-mode operation. This aspect will be further investigated in the future.}, and despite the variation of ellipticity over the 2:1 factor band, the maximum limits are acceptable.

\begin{table*}
\centering
\caption{Variation with frequency of the main beam characteristics in the operational bandwidth (10--20\,GHz), for the 0\textdegree~polarization.}
\label{tab:beamvsfreq_0pol}
\resizebox{0.7\textwidth}{!}{%
\begin{tabular}{ccccccc}
\hline
\begin{tabular}[c]{@{}c@{}}Frequency \\ (GHz)\end{tabular} & \begin{tabular}[c]{@{}c@{}}Edge Taper\\ (dB)\end{tabular} & \begin{tabular}[c]{@{}c@{}}$\mathrm{D_{feed}}$\\ (dBi)\end{tabular} & \begin{tabular}[c]{@{}c@{}}$\mathrm{D_{refl}}$\\ (dBi)\end{tabular} & \begin{tabular}[c]{@{}c@{}}FWHM\\ (arcmin)\end{tabular} & \begin{tabular}[c]{@{}c@{}}$e$\\ (\%)\end{tabular} & \begin{tabular}[c]{@{}c@{}}XPD\\ (dB)\end{tabular} \\ \hline \hline
10.0 &$-$18.30  &18.63  &37.48  &126.6  &0.66  &$-$30.50    \\
10.5 &$-$19.50  &18.98  &38.24  & 121.5 &0.17  &$-$31.13    \\
11.0 &$-$20.36  &19.34  &38.76  &116.6 &0.41  &$-$32.46   \\
11.5 &$-$20.97  &19.70  &39.19  &112.7  &0.81  &$-$33.75   \\
12.0 &$-$21.55  &20.06  &39.62  & 109.5  & 0.12  &$-$34.37   \\
12.5 &$-$22.16  &20.42  &39.90  &106.7  &0.36  &$-$34.87    \\
13.0 &$-$23.44  &20.79  &40.13  &104.4  &0.32  &$-$34.82   \\
13.5 &$-$24.30  &21.06  &40.39  &   102.5  &0.53  &$-$34.94    \\
14.0 &$-$26.74  &21.33  &40.57  & 100.8    & 0.67  &$-$34.90   \\
14.5 &$-$30.89  &21.62  &40.72  &99.3 &0.90  &$-$36.35   \\
15.0 &$-$33.12  &21.91  &40.83  &98.5  &1.59  &$-$37.32   \\
15.5 &$-$34.86  &22.19  &40.93  &97.9  &2.18  &$-$36.11   \\
16.0 &$-$34.56  &22.46  &40.97  &97.8  &3.16  &$-$36.87   \\
16.5 &$-$33.99  &22.58  &40.99  &97.8   &4.09  &$-$36.64    \\
17.0 &$-$31.89  &22.69  &41.00  &98.2  &4.76  &$-$35.63   \\
17.5 &$-$31.11  &23.0   &41.01  &98.7   &5.02  &$-$36.40   \\
18.0 &$-$30.36  &23.32  &40.96  &99.2   &5.24  &$-$36.58   \\
18.5 &$-$28.54  &23.46  &40.92  &100.0  &5.85  &$-$36.65  \\
19.0 &$-$31.52  &23.59  &40.84  &101.4 &7.10  &$-$36.48   \\
19.5 &$-$32.04  &23.56  &41.30  &95.7  &7.82  & $-$37.19   \\
20.0 &$-$37.29  &23.54  &41.00  &98.2 &1.04  &$-$38.72  \\ \hline
\end{tabular}%
}
\end{table*}

\begin{table*}
\centering
\caption{Variation with frequency of the main beam characteristics in the operational bandwidth (10--20\,GHz), for the 90\textdegree~polarization.}
\label{tab:beamvsfreq_90pol}
\resizebox{0.7\textwidth}{!}{%
\begin{tabular}{ccccccc}
\hline
\begin{tabular}[c]{@{}c@{}}Frequency \\ (GHz)\end{tabular} & \begin{tabular}[c]{@{}c@{}}Edge Taper\\ (dB)\end{tabular} & \begin{tabular}[c]{@{}c@{}}$\mathrm{D_{feed}}$\\ (dBi)\end{tabular} & \begin{tabular}[c]{@{}c@{}}$\mathrm{D_{refl}}$\\ (dBi)\end{tabular} & \begin{tabular}[c]{@{}c@{}}FWHM\\ (arcmin)\end{tabular} & \begin{tabular}[c]{@{}c@{}}$e$\\ (\%)\end{tabular} & \begin{tabular}[c]{@{}c@{}}XPD\\ (dB)\end{tabular}  \\ \hline \hline
10.0 &$-$18.30  &18.63  &37.27  &126.6  &0.71  &$-$31.16    \\
10.5 &$-$19.50  &18.98  &38.16 & 121.6 &0.22  &$-$32.88    \\
11.0 &$-$20.36  &19.34  &38.71  &116.7 &0.34  &$-$33.51   \\
11.5 &$-$20.97  &19.70  &39.17  &112.8  &0.27  &$-$33.28   \\
12.0 &$-$21.55  &20.06  &39.58  &109.7  &0.16  &$-$35.25   \\
12.5 &$-$22.16  &20.42  &39.85  &106.9  &0.25  &$-$36.51    \\
13.0 &$-$23.44  &20.79  &40.09  &104.5  &0.70  &$-$37.70   \\
13.5 &$-$24.30  &21.06  &40.34  &102.5  &0.83  &$-$37.70    \\
14.0 &$-$26.74  &21.33  &40.53  &100.9 &1.44  &$-$37.42   \\
14.5 &$-$30.89  &21.62  &40.68  &99.4  &1.47  &$-$36.80   \\
15.0 &$-$33.12  &21.91  &40.81  &98.5  &2.92  &$-$39.93   \\
15.5 &$-$34.86  &22.19  &40.91  &98.0  &3.23  &$-$36.79   \\
16.0 &$-$34.56  &22.46  &40.94  &97.9  &4.17  &$-$37.43   \\
16.5 &$-$33.99  &22.58  &40.97  &97.9  &4.92  &$-$37.46    \\
17.0 &$-$31.89  &22.69  &40.98  & 98.3  &5.87  &$-$37.73   \\
17.5 &$-$31.11  &23.0   &40.99  &98.6  &5.85  &$-$37.51   \\
18.0 &$-$30.36  &23.32  &40.96  &99.1  &6.19  &$-$37.69   \\
18.5 &$-$28.54  &23.46  &40.92  &99.8  &6.65  &$-$38.07  \\
19.0 &$-$31.52  &23.59  &40.81  &101.6  &8.32  &$-$38.02   \\
19.5 &$-$32.04  &23.56  &41.3  &95.4  &7.51  & $-$38.10   \\
20.0 &$-$37.29  &23.54  &41.00  &97.9  &1.65  &$-$38.49  \\ \hline
\end{tabular}%
}
\end{table*}

\subsection{Far sidelobes characterization} \label{sec:farlobes}

Here we analyse of the $\mathrm{4\pi}$ beams, taking into account not only the feedhorn and mirror systems, but also assessing the need for a shielding structure. Local maxima in the radiation pattern that are not part of the main beam, are dominated by diffraction effects. These maxima, or sidelobes, must be carefully analysed in order to properly mitigate them, since they can introduce an unwanted contribution to the sky temperature. This measurement contamination depend on several factors, namely the observed sky region (in the farlobes angular range), the frequency band and the shielding --- if necessary --- and the side lobe  efficiency.

\begin{figure}
    \centering
    \subfloat{\includegraphics[width=0.4\textwidth]{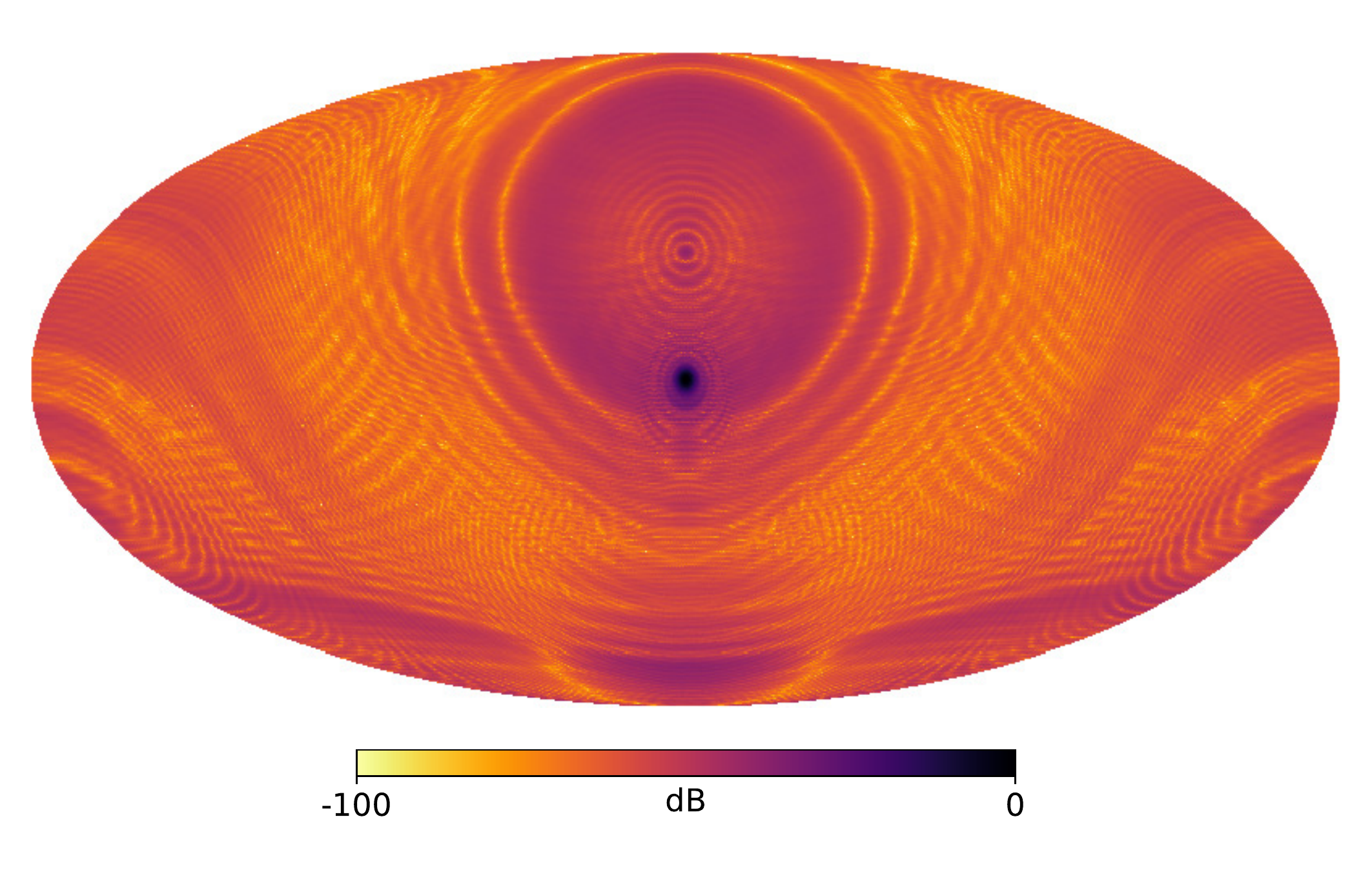}\label{subfig:pyrbed}}
    \qquad
    \subfloat{\includegraphics[width=0.4\textwidth]{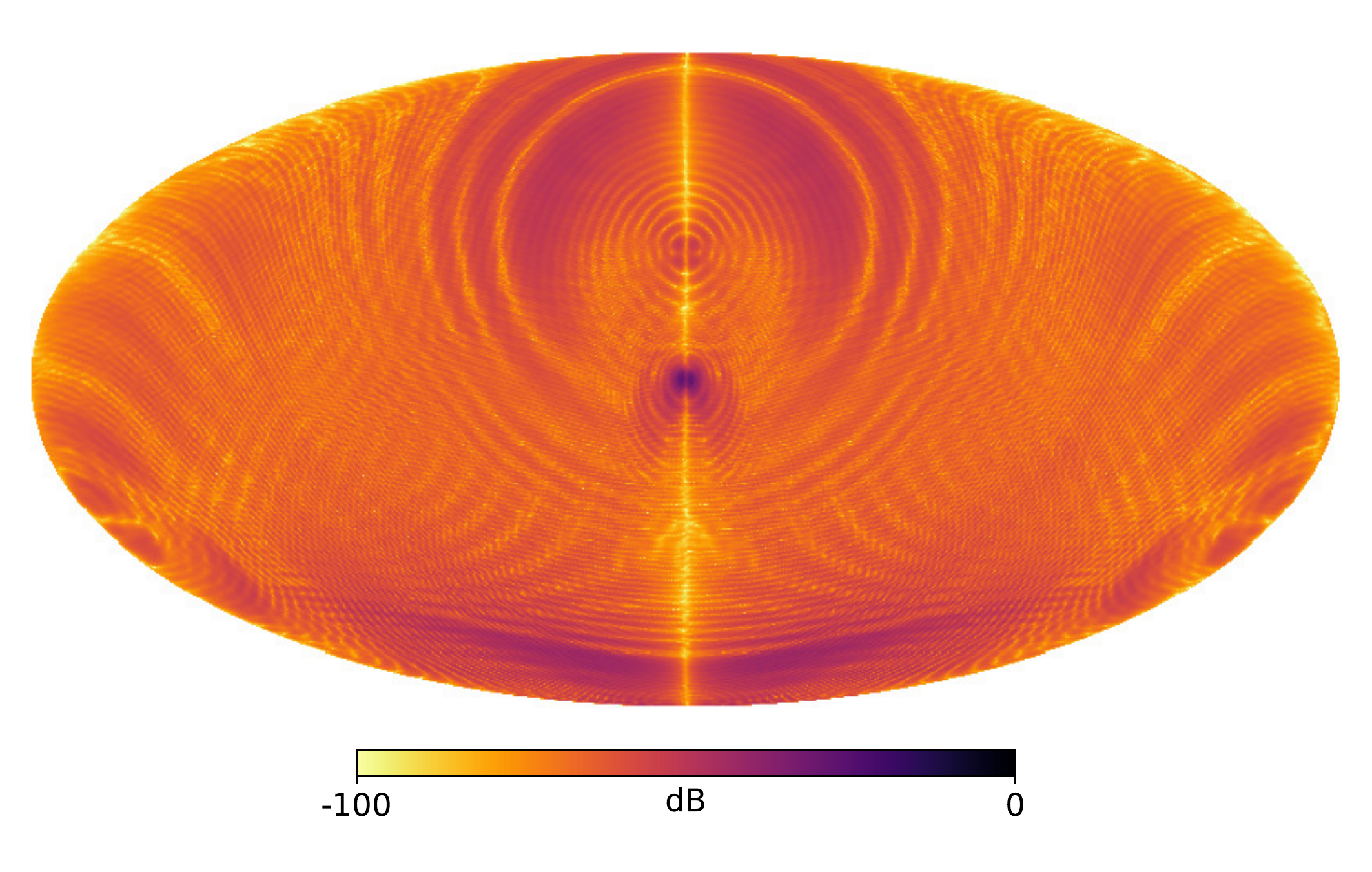}\label{subfig:pyrbed}}
    \\ 
    \vspace{-5mm}
    \subfloat{\includegraphics[width=0.4\textwidth]{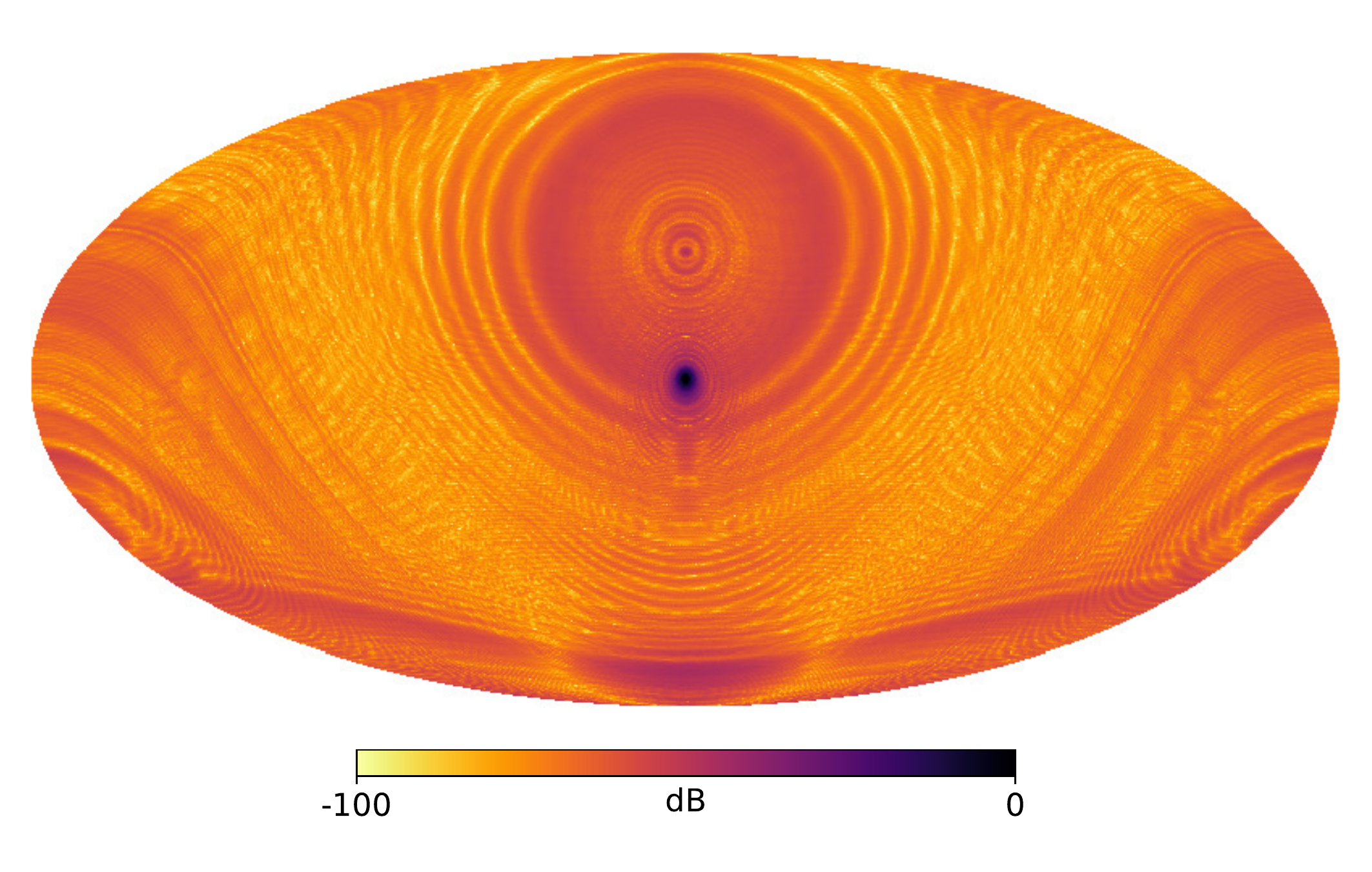}\label{subfig:pyrbed}}
    \qquad
    \subfloat{\includegraphics[width=0.4\textwidth]{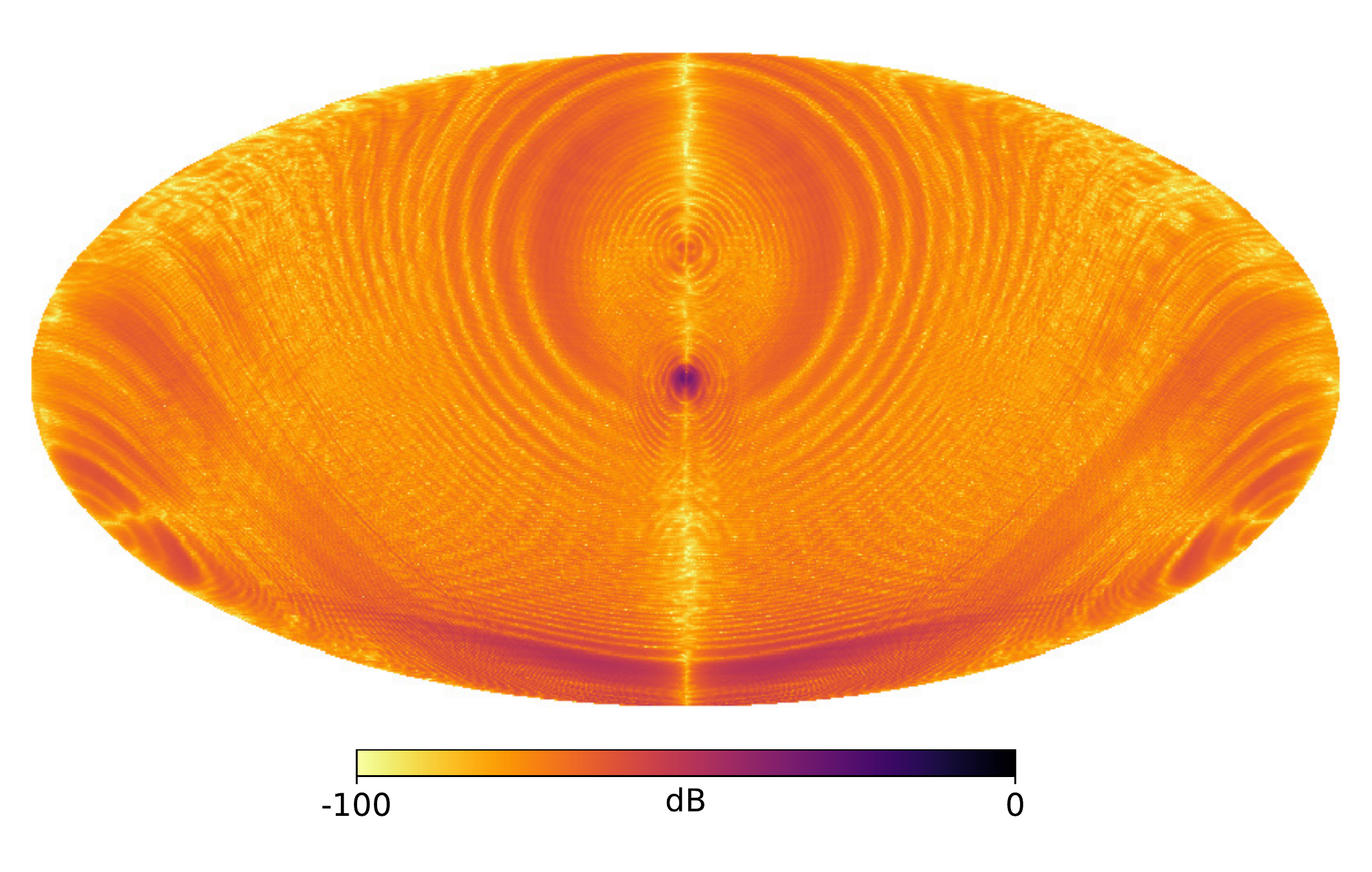}\label{subfig:pyrbed}}
    \\
    \vspace{-5mm}
    \subfloat{\includegraphics[width=0.4\textwidth]{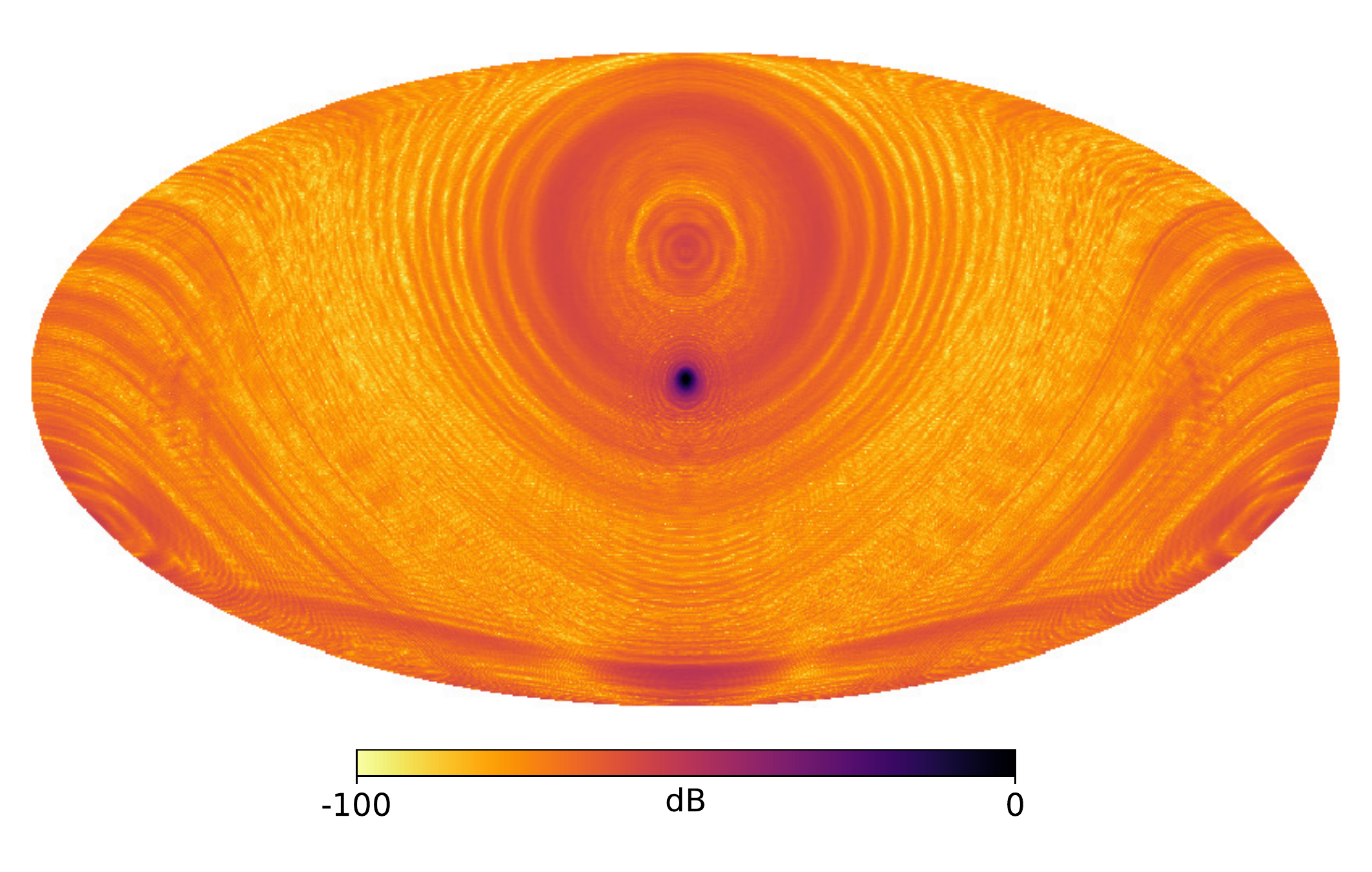}\label{subfig:pyrbed}}
    \qquad
    \subfloat{\includegraphics[width=0.4\textwidth]{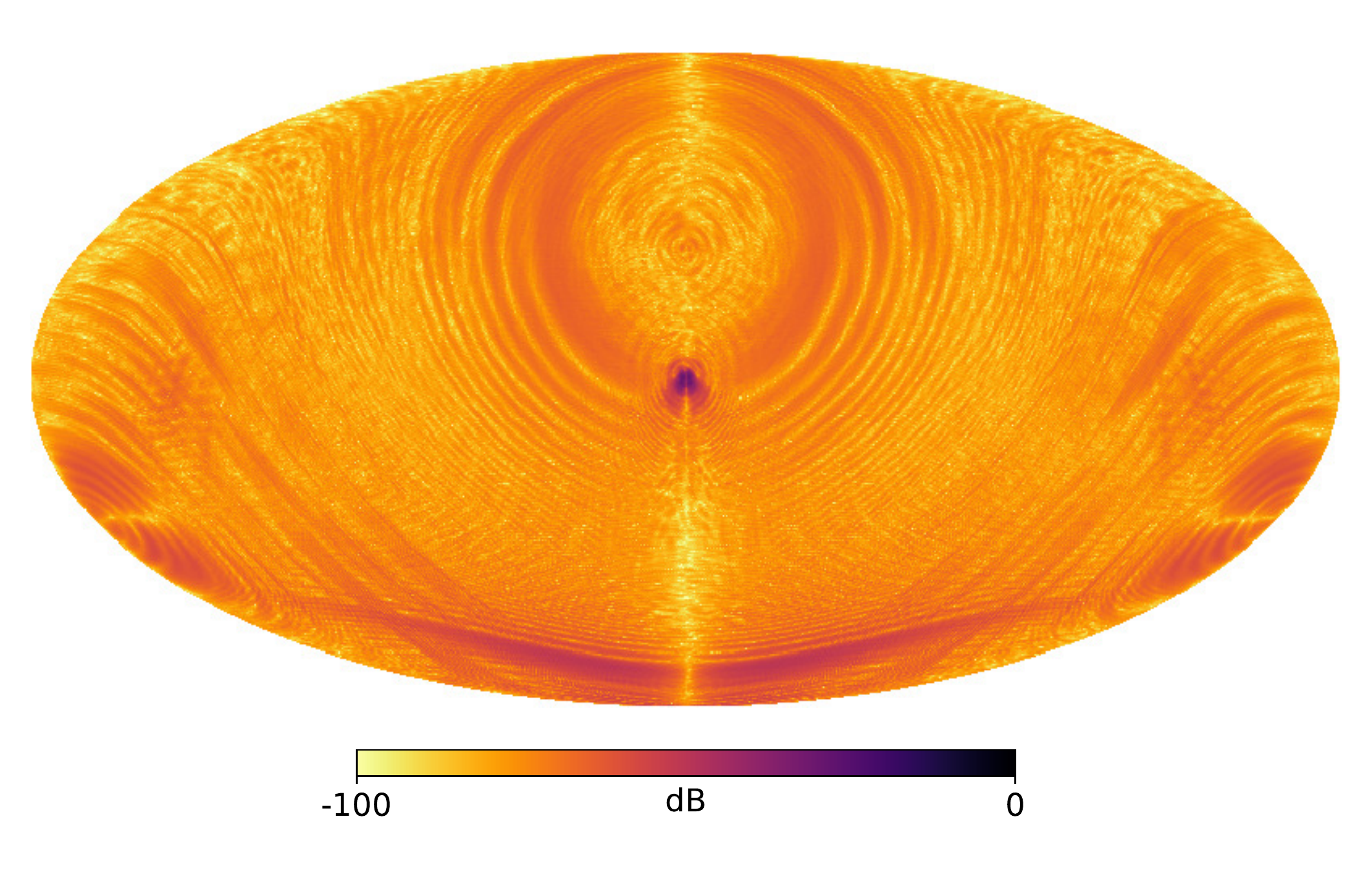}\label{subfig:pyrbed}}
    \\
    \caption{Radiation pattern over the full sky computed and its variation with frequency in the TMS range. From \emph{top} to \emph{bottom}, we include simulations at 10, 15 and 20\,GHz.  \emph{Left}: Copolar components. \emph{Right:} cross-polar component. To facilitate direct comparison, the patterns have been normalised with respect to the directivity value in the direction of maximum radiation in each sub-band, shown in table~\ref{tab:beamvsfreq_0pol}.} 
    \label{fig:fullsky}
\end{figure}

Mollweide projections of the co-polar and cross-polar components of the radiation pattern are shown in figure~\ref{fig:fullsky}, for 10, 15 and 20\,GHz. These maps have been projected to the sphere using HEALPix routines\footnote{\url{https://healpix.sourceforge.io}}. The effect of diffraction at the edges of the secondary mirror is the most striking and critical parameter, as it generates lobes well beyond the requirements of the system. Mitigation of these lobes can be achieved by shaping the edges of the mirror, shielding the back of the mirror with a metallic plate coated with absorbent material, or both.  These actions would allow us to reduce  lobe levels by about $-$15\,dB, based on the QUIJOTE results and the specifications for Eccosorb ANW-77  from 10--20\,GHz, \cite{eccosorb}.

\begin{figure}
    \centering
     \subfloat{\includegraphics[width=0.45\textwidth]{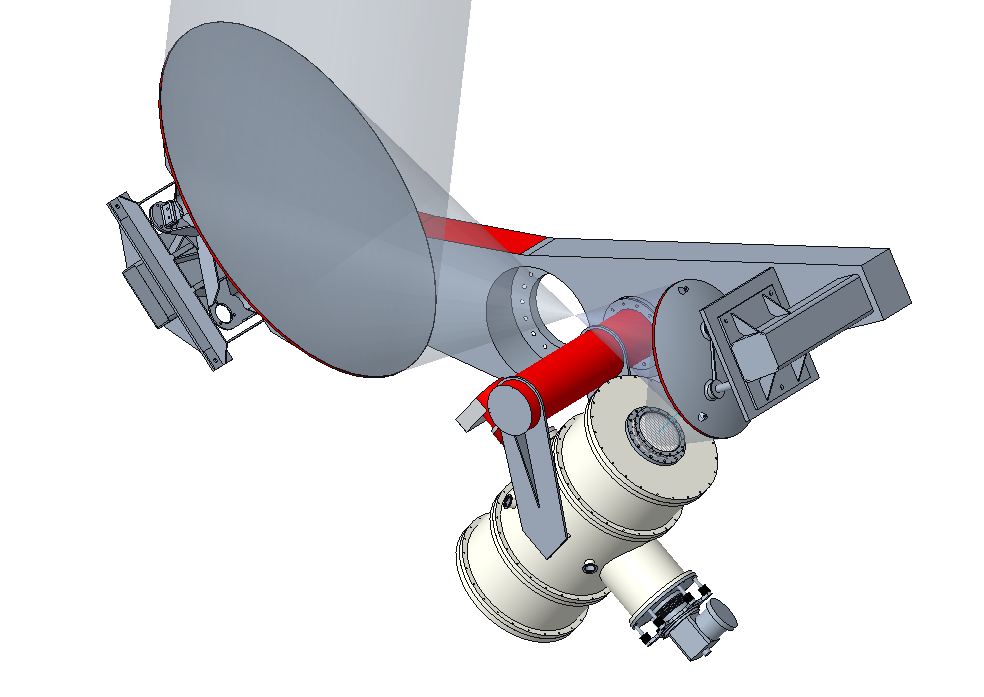}\label{subfig:support}}
    \qquad
    \subfloat{\includegraphics[width=0.45\textwidth]{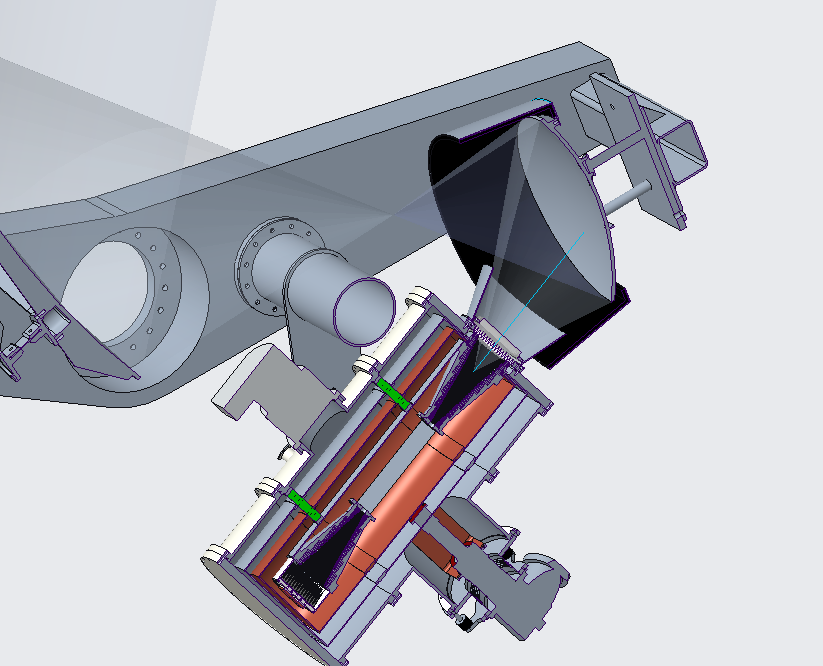}\label{subfig:shielding}}
    \caption{Mechanical design of the optical system, supporting structure and proposed shielding structure. \emph{Left:} Designed structure to support the TMS dual-reflector system and the cryostat containing the TMS instrument. Highlighted in red colour are the parts most likely to compromise the performance of the system in terms of sidelobes. \emph{Right:} Co-moving shielding structure proposed to reduce  sidelobes due to diffraction on the SM edges.  The aluminum plates will be covered with an absorber material, typically Eccosorb. We are currently evaluating a complete co-moving structure surrounding the optical system with a single circular aperture for the main beam entrance. }
    \label{fig:structure}
\end{figure}

We have highlighted in figure~\ref{subfig:support} the critical points of the system, i.e. the mirror edges, the cryostat pump and nearby supporting structures, where diffraction and scattering phenomena will take place.  The design of the optical system  already partially reduces these effects by providing a taper illumination of at least $-$20\,dB, as explained in section~\ref{sec:edge}. In addition, we propose a shielding structure that co-moves with the dual-reflector system, in such a way as to prevent unwanted radiation, as figure~\ref{subfig:shielding} illustrates. The shielding metal plates are to be covered by a layer of absorbent material (eccosorb), with a thickness of at least 20\,mm. In this study, no simulations have been carried out to verify the effectiveness of the proposed shielding due to the limited computing resources. We therefore leave this task pending for the future.

\begin{figure}
    \centering
    \includegraphics[width=0.7\textwidth]{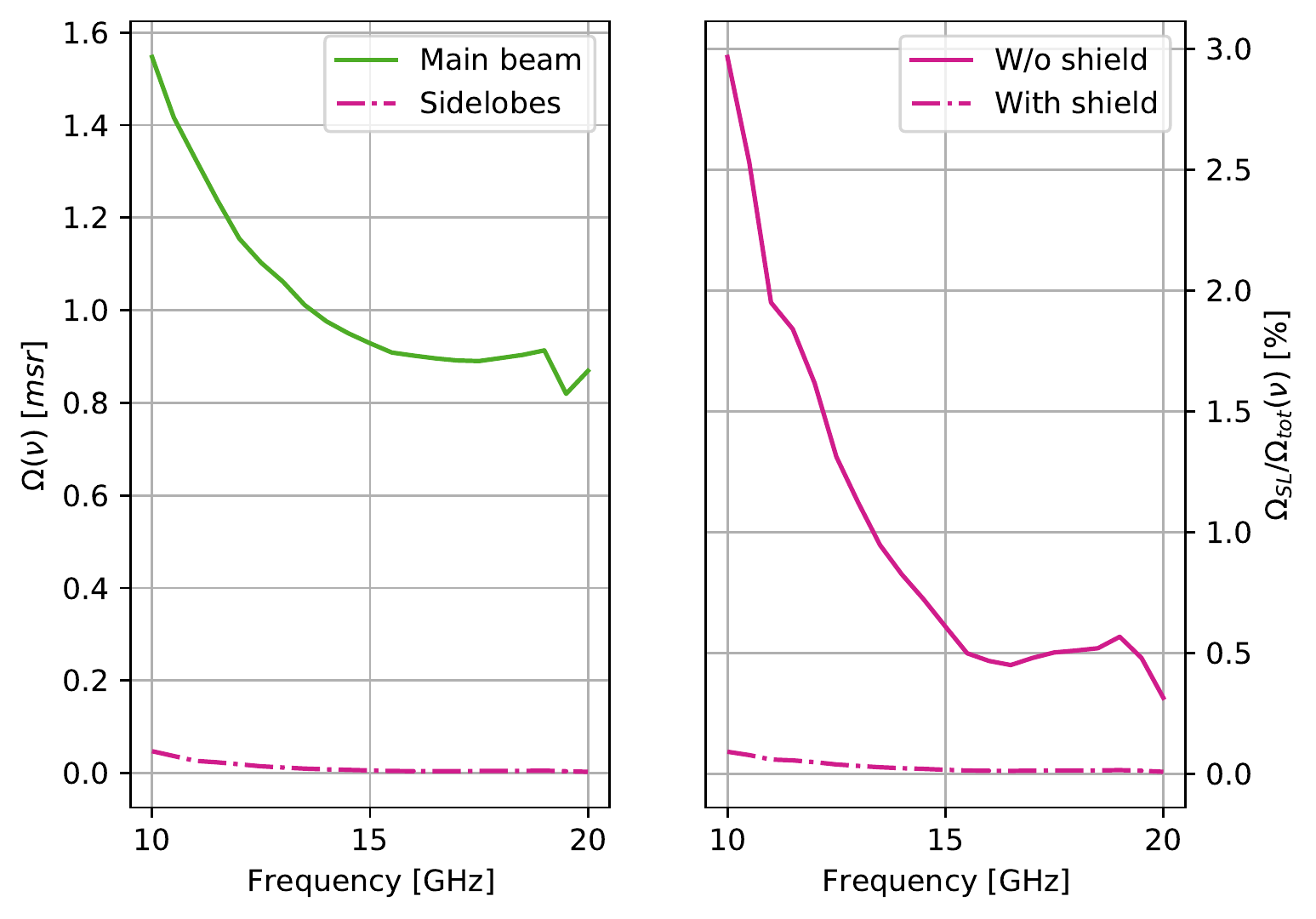}
     \caption{Beam solid angle frequency variation computed from the directivity angular distribution obtained with CST simulations.  \emph{Left}: Comparison of the solid angle corresponding to the sidelobes, shown in \emph{pink},   versus the main beam, associated with the \emph{green colour},  angular region, for the optical system without a shielding structure. \emph{Right}: Relative contribution of the sidelobes, and its variation within the frequency band. \emph{Solid} line corresponds to the original ratio between the solid angle associated to the sidelobes region versus the total solid angle. With \emph{dashed} line, we show a prediction for the sidelobes contribution with the proposed shielding structure, which we expect to reduce below 0.5\,\%. }
    \label{fig:solidangle}
\end{figure}

In order to conclude the frequency characterisation of the beam, we have compared the relative contribution of the main beam with respect to that of the sidelobes. We do so by means of the solid angle, which we calculate  by approximating the integral of the directivity angular distribution as

\begin{equation}
\Omega_{ML, SL} =   \oint_{ML,SL~ reg.}{D\left(\theta,\phi \right ) d\Omega}~\approx \sum_{ML,SL~reg._i}{D\left(\theta_i,\phi_i \right )\sin(\theta_i)~\Delta\theta_i \Delta\phi_i}   
\end{equation}

and defining the main lobe (ML) region as the angular region containing the main lobe up to the first minimum, and the sidelobes (SL) region as the remaining angular positions. Figure~\ref{fig:solidangle} illustrates the variation on the solid angle in the TMS band. We have represented the variation in frequency of the solid angle, corresponding to both the ML and SL region. 

We observe a 40\% variation of the main beam solid angle between the band limits, i.e. between 10 and 20\,GHz. We emphasise the importance of correcting this effect in the calibration software, as it is by no means negligible. On the other hand, the ratio between sidelobes solid angle versus total solid angle has a peak value of approx.\,3\% at the lowest edge of the band (10\,GHz). Based on the experience gained with the QUIJOTE experiment, we expect to further mitigate this contribution by means of the proposed shielding below relative levels of 0.1\%.

\subsection{Polarization leakage} \label{sec:polleak}

Finally, another crucial instrumental effect that needs to be fully characterised and controlled is polarization leakage. As anticipated in section~\ref{sec:overview}, TMS will have polarimetric capabilities, and thus, it will measure the I-Q-U  spectra. However, we must consider systematic effects such as the instrumental coupling between the Stokes I and the linearly polarized Q-U parameters. The Müller matrix, which describes these couplings, depends on the telescope beam. 

Following a similar formalism used by \cite{leahy2010}, we can calculate the beam Müller matrix separatedly for every frequency in the TMS band, from the Jones matrix elements. In the case of the optical system, the Jones matrix is

\begin{equation}
\label{eq:jonesmat}
   \mathcal{J}^{beam} = \frac{1}{\sqrt{4\pi}} \begin{pmatrix}
   \textrm{Copolar}_{\rho=0^\circ} &\textrm{Xpolar}_{\rho=0^\circ}\\
    \textrm{Xpolar}_{\rho=90^\circ} &\textrm{Copolar}_{\rho=90^\circ}
\end{pmatrix}
\end{equation}

where the elements of this matrix are associated with the copolar and cross-polar components for each linear polarizations, 0\textdegree~and 90\textdegree. In fact, these elements correspond to the radiation patterns computed in previous sections. We thus calculate the angular distribution of the Müller matrix elements, i.e., each element of the matrix is like a radiation pattern in itself. Assuming that the cross terms are small enough to be ignored, ---and we are talking about a ratio of less than 2.8\% in the worst cases (10\, GHz) with respect to the auto-correlation terms---, we can simplify the Müller matrix expression considerably. In addition, the diagonal components are almost identical, and the same occurs for the antidiagonal elements, resulting in the matrix as follows
 
\begin{eqnarray}
\label{eq:mullermat}
   \nonumber \mathcal{M}_{II} \approx \mathcal{M}_{QQ} \approx \frac{1}{2} \cdot \left ( \mathcal{J}_{11} + \mathcal{J}_{22} \right ) \\
    \mathcal{M}_{IQ} \approx \mathcal{M}_{QI} \approx \frac{1}{2} \cdot \left ( \mathcal{J}_{11} - \mathcal{J}_{22} \right ) 
\end{eqnarray}

\begin{figure}
    \centering
    \subfloat{\includegraphics[width=0.35\textwidth]{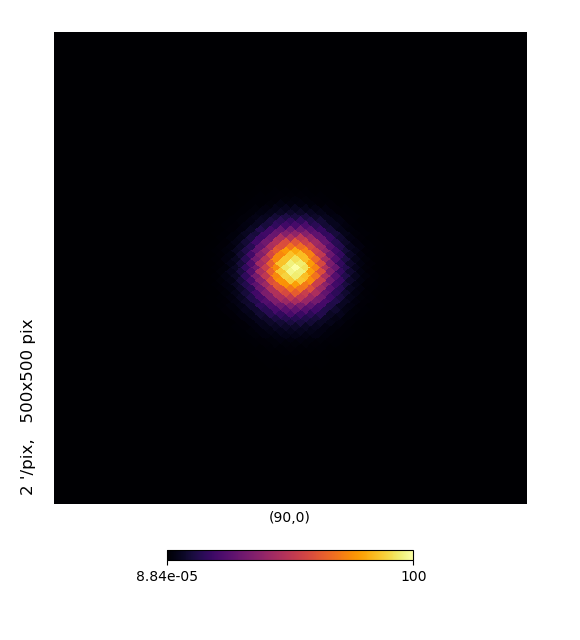}}
    \qquad
    \subfloat{\includegraphics[width=0.35\textwidth]{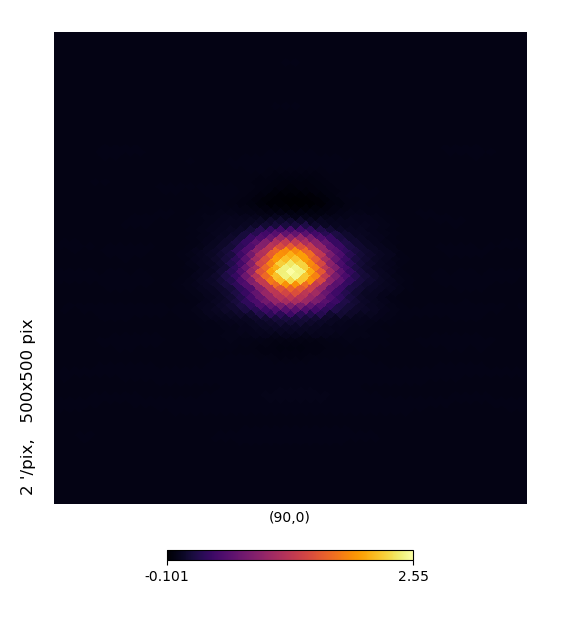}}\\ \vspace{-8mm}
    \subfloat{\includegraphics[width=0.35\textwidth]{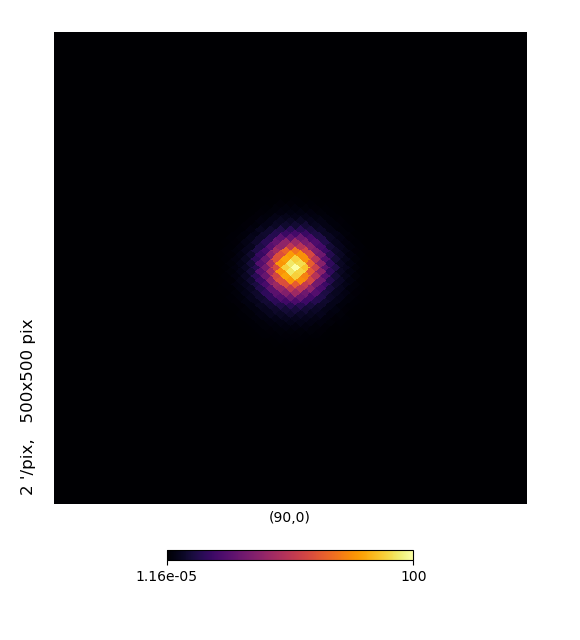}}
    \qquad
    \subfloat{\includegraphics[width=0.35\textwidth]{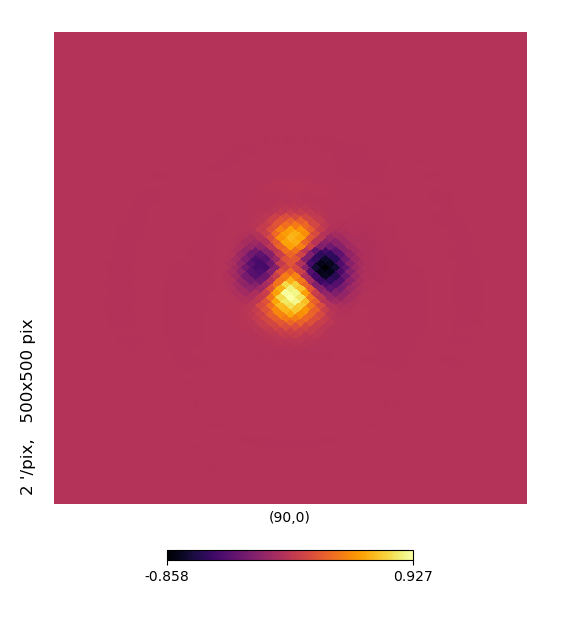}}\\ \vspace{-8mm}
        \subfloat{\includegraphics[width=0.35\textwidth]{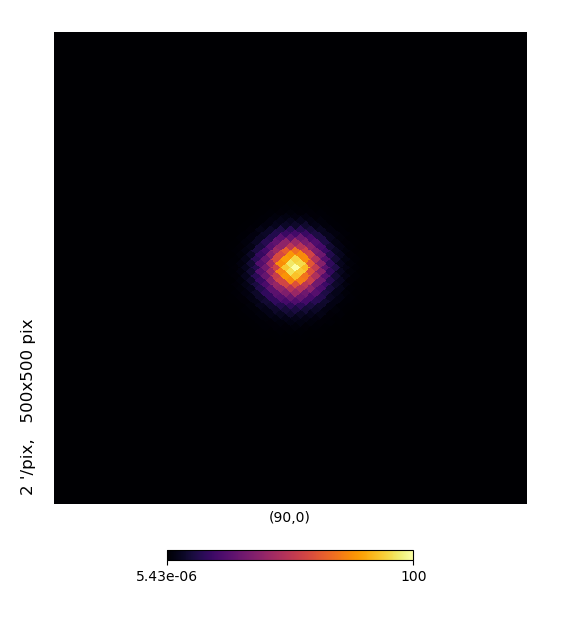}}
    \qquad
    \subfloat{\includegraphics[width=0.35\textwidth]{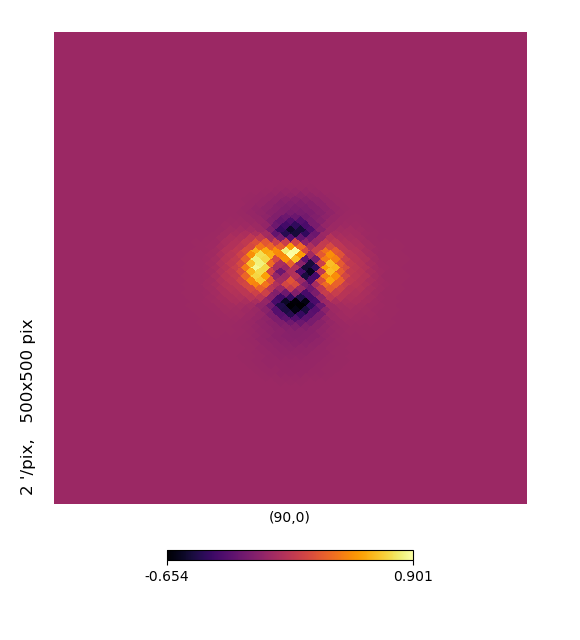}}\\
    \caption{Beam Müller matrix representing the effect of the TMS optical system on the Stokes parameters at, from \emph{top} to \emph{bottom,} 10, 15 and 20\,GHz. All beam patterns have been normalised and expressed as percentages, in such a way that the directivity value of the main beam corresponds to 100\,\%. \emph{Left:} Symmetric matrix elements $\mathrm{M_{II}}$ and $\mathrm{M_{QQ}}$. \emph{Right:} Matrix element $\mathrm{M_{IQ}}$.}
    \label{fig:beammuller}
\end{figure}

The TMS main beam system presents a slight ellipticity, dependent also on the frequency band and the polarization, as shown in tables~\ref{tab:beamvsfreq_0pol}~and~\ref{tab:beamvsfreq_90pol}. This difference in beam shape and size for the two orthogonal polarization causes the so-called effect of \emph{beam squash}, \cite{robishaw2018}. Beam squash has a direct impact on the $\mathrm{\mathcal{M}_{IQ}}$ and $\mathrm{\mathcal{M}_{QI}}$ patterns, introducing in most bands a four-lobed cloverleaf pattern, with two positive lobes facing each other, and two negative lobes with a rotation of 90\textdegree. Figure~\ref{fig:beammuller} presents the beam Müller matrix, calculated used the approximation in eq.~(\ref{eq:mullermat}), where we can see the evolution of the cross-term patterns at 10, 15 and 20\,GHz. We expect a maximum excess in the absolute temperature of about 2.55\% and 0.90\%, at 10 and 20\,GHz respectively, when reconstructing the spectrum of the Q-U Stokes parameters due to polarization leakage.

\section{Conclusions}\label{sec:conclusions}

The response of the optical system is a major contributor in any astronomical observation, and unavoidably impacts on sky measurements. Non-negligible sidelobes, the main beam shape --- its ellipticity ---, and the introduction of a cross-polarized component introduce spurious effects and must be controlled. In the specific case of the TMS being a spectrometer, a special effort must be made to understand the variation within the band.

The present work provides a full description of the optical TMS system and of its constituent parts, 
and a detailed description of the method and rationale behind the design decisions. We present also the first comprehensive assessment of the response of the TMS optical system. We have performed detailed electromagnetic simulations to determine the optical response, identify critical points of the design and observe whether the requirements are met.

For the simulations we have included the TMS feedhorn, vacuum window and cryostat, and the dual-reflector system. We have studied the effect of the UHMWPE vacuum window and the IR filter on the radiation pattern of the TMS feedhorn. We have performed an in-deph analysis of the radiation pattern of the complete optical system, both of the main beam --- its shape and size, but also the cross-polarization contribution ----, and the sidelobe region. We have placed special emphasis on the variation of the different figures of merit within the TMS band between 10 and 20\,GHz, including the variation in each frequency band of the sidelobes contribution with respect to the main beam. Finally, we have analysed the polar response of the optics, which will also introduce some errors in the reconstruction of the Stoke parameters spectra.

Being limited by the computational and software resources, this study did not evaluate the efficiency of the proposed shielding structures in mitigating the far sidelobes, especially those due to diffraction phenomena on the secondary mirror rim. There has also been no modelling of possible manufacturing imperfections --- e.g. irregularities on mirror surfaces, de-focusing and horn tilting, among others. Further modelling work will be conducted in order to obtain a better understanding of this system.

In spite of the limitations of the study, the understanding we have gained of the radiation patterns will be a fundamental tool in the subsequent calibration and scientific exploitation phases, as it will facilitate the reconstruction of the real sky spectrum.

\FloatBarrier

%%%%%%%%%%%%%%%%%%%%%%%%%%%%%%%%%%%%%%%%%%%%%%%%%%%%%%
\acknowledgments

This work has been partially funded by the Spanish Ministry of Science and Innovation (MICINN) under the projects IACA15-BE-3707, EQC2018-004918-P, PID2020-120514GB-I00 and the Severo Ochoa Programs SEV-2015-0548 and CEX2019-000920-S. We also acknowledge financial support from the Spanish MICINN under the FEDER Agreement INSIDE-OOCC (ICTS-2019-03-IAC-12).

Some of the results in this paper have been derived using the HEALPix package \cite{gorski2005}.  The authors would like to acknowledge the CST Studio Suite for use of their 3D EM simulation and analysis software, and the Student Edition of the GRASP tool from TICRA used in the preliminary analysis described in this work.  

We acknowledge the support of all the technicians, engineers, scientists and administrative staff of the IAC and QUIJOTE experiment.

\bibliographystyle{unsrt}
\bibliography{biblio}

\end{document}